\documentclass[12pt]{article}
\usepackage{graphicx}
\usepackage{amssymb}
\usepackage{epstopdf}
\date{}

\title{Formaleuros, Formalbitcoins, and Virtual Monies}
\author{
	Jan A. Bergstra
	 \\
\\
  {\small
	  Informatics Institute, Faculty of Science, 
	  University of Amsterdam}\\
	{\small Email: \texttt{j.a.bergstra@uva.nl}}
	}

\begin{document}
\bibliographystyle{plain}
\maketitle
\begin{abstract}
Formalist positions towards money are considered from a perspective of formal methods in computing. The Formaleuro 
(FEUR) as a dimension
for monetary quantities is proposed as well as the Formalbitcoin (FBTC) which represents an 
item ready for circulation in a model of informational money. The rationale of these notions is illustrated though formulating
questions about monies in a terminology that profits from this indirection. 

An attempt is made to understand the concept of money from scratch given the wider context of existing scientific and 
philosophical work on money and finance. The sheer size
of that literature and the seemingly hopeless task to find out what has already been done is taken as an incentive to 
analyze in rather unusual detail
how to get started in a subject where the plan to develop a significant knowledge of existing work may be unfeasible.

In order to provide a definition of money the need is felt to make use of a tailored theory of definition. To that end a 
theory of imaginative definitions is presented and its implications for definitions of money are sketched.

It is argued that a theory of money may be dependent  on the role of its holder. A survey of some roles is given,
with the so-called subordinate administrative role (SAR)  in a central position. 
A specialized but informal theory of money is proposed for  the subordinate administrative role.

The concepts of virtual memory and virtual machine are taken as the point of departure for a definition of
the notion of virtual money. 
It is argued that from the perspective of a component (division) 
of a large organization (ORG) its local financial system (LFS) provides 
a virtual money vm(LFS, ORG) which may well fail to meet the 
most common general and acknowledged moneyness criteria. Inverse
moneyness preference is coined as phrase to assert the tendency of top-management of 
ORG to make its virtual money deviate from these criteria.
\end{abstract}

\newpage\tableofcontents\newpage

\section{Informaticology of money and finance (IoM\&F)}\label{SecTIMF}
This paper\footnote{%
Below ``he or she'' will be abbreviated to ``he''. Similar abbreviations have been applied throughout the text.
This paper is a revised version of our ``Formaleuros, formalcoins, and virtual monies'' (\texttt{arXiv.org/abs/1008.0616v1})
written in 2010.
The revision consists of (i) the removal of (too many) typo's, (ii) bringing the paper up to date with the appearance of Bitcoin (and including
some results from recent joint work on Bitcoin with Karl de Leeuw in \cite{BergstraL2013}),
(iii) a significant update of the section on Islamic finance and interest prohibition 
(drawing on recent joint work with Kees Middelburg in \cite{BM2011}, as well as on some remarks made in \cite{BergstraL2013}), 
(iv) deletion of some 4 pages in total of fragments of minor importance, (v) a minor update of references,
(vi) moving the ad hoc theory of definitions to an appendix, and (vii) a small change of the title.}
is about money and finance in quite general terms. We intend to position this paper as 
a piece of work in theoretical informatics\footnote{%
Following \cite{Bergstra2012d} we prefer to use the term informaticology instead of
theoretical informatics; informaticology having a different scope, 
less mathematical and with some proximity to philosophy of science. 
An alternative phrase for naming IoM\&F is ``informaticological finance''.} 
rather as work in economy or any other social science. 
Some reasons for doing so are:
\begin{enumerate}
\item We  expect that a number of techniques that have been developed in 
theoretical informatics can be meaningfully applied to money and finance.
\item In particular we expect this to be the case for a number of techniques 
from the semantic part of theoretical computer science, including fairly standard 
techniques that we have been using for many years, such as 
process algebra, module algebra and equational abstract data type specifications, 
and much less known recent developments such 
as thread algebra \cite{BergstraMiddelburg2007}), interface groups (\cite{BergstraMiddelburg2010b}), 
tuplix calculus (\cite{BPZ2007,BNZ2008}), promise theory (\cite{Burgess2007}),
and meadows (\cite{BT2007,BBP2013}).
\item We assume that it is just a matter of time that all manifestations of money are 
exclusively based on information technology.
Then money will be a major application of computer science and the theory of 
money will be a part of theoretical informatics. 

Moini \cite{Moini2001} extensively discusses the increasing dependence of money on information technology. 
Moini concludes that money cannot be a thing, it is a credit debt relation between different agents 
(essentially an informational item but 
perhaps represented by a thing) 
and monetary systems have always been information systems.
\item For communities of human, artificial, and corporate agents, 
money is a data structure equipped with a number of protocols which supports  
the cooperation of agents each proceeding towards their particular objectives. 
Money constitutes part of the world wide operating system 
so to speak and from a computing perspective its role is intriguing just because it 
supports the concurrent activity of a large number of heterogeneous
agents in an effective fashion. 
\item In other words, and using computer science terminology: 
money is a concurrency control data type comparable to Dijkstra's 
semaphores and Hoare's monitors.
\item It is conceivable, though not  investigated below, that concurrency control of 
artificial agents can also be brought forward by the 
introduction of some form of money. That perspective requires 
a semantic understanding of money from a perspective of informatics, and in particular from the area of program notation semantics.
\item Money, as a formal concept brought to bearing back into informatics, 
can be considered a mechanism for defining aggregate complexity 
measures which can incorporate many disparate cost factors. 
This may be needed for explaining or analyzing the utility of methods and techniques in computing which
fail to deliver  a tangible advantage from the perspective of one of the known complexity 
measures but which nevertheless provide an 
advantage at some aggregate level. To analyze cost advantages at the aggregate level, 
different and in principle incomparable cost
factors need to be simultaneously accommodated,  combined and compared in combined form. Money based cost analysis 
provides a way to linearize combined cost factors. Economics provides a tradition of balancing disparate utilities in spite of
the absence of a formal and fully reliable methodology for comparing different utilities beforehand.
\item Money and finance give rise to decisions and play an important role in making decisions. 
For instance Kirchler \cite{Kirchler1995}
writes that financial decisions in housekeeping need to be considered a part of concurrent activity, 
whereas decisions that are routinely taken
make use of so-called sequential programs. Connections with the widespread usage of concurrent and sequential programs
in computing seem to have been ignored. But it is quite clear that decision making processes may 
run in parallel and that multi-threading\footnote{%
We refer to \cite{BergstraMiddelburg2007} for a formalization of multi-threading which we expect 
to be be useful for the  formalization of decision making.} may be useful for the description of this kind of concurrency.
\end{enumerate}

\subsection{Formal methods, a generic part of theory}
In appendix A we have listed some of our own work which we consider potentially useful 
for work on money from a background in theoretical informatics. 
From that work we have concluded that there is a sufficient justification for  investigating money from the 
point of view of theoretical 
informatics and in particular for research on money and finance that makes use of the various tools that belong to  the semantic branch 
of theoretical informatics which the author 
has been using for a number of years. Like in computing we assume that a formal methods area can be outlined within the larger body of theory, 
the formal methods being mainly concerned with logic, semantics, reasoning and formalisms, while theory at large includes algorithms, complexity,
probabilistic methods and simulation techniques.

\subsubsection{Ontology needed but missing}
A stable and generally accepted ontology providing a workable map of terms and notions 
concerning what we will call money items seems to be missing. Here is an attempt: 
\begin{enumerate}
\item Coin and Banknote (the capital indicative of a class rather than an instance of it), are subclasses of Currency (in the narrow sense
of just meaning an abstract superclass of those two kinds of items).
\item A coinage (also called a coin family) provides a coherent collection of coin types, 
moving through a steady evolution.\footnote{%
We consider it plausible to conjecture that 25 years from now coinage (and banknote families) 
will not be included in the then valid moneyages anymore.}
 Different coinages are usually
disjoint with respect to their coin types.
\item A coin type type usually consists of a series of coin type versions, each coin type version having instances (true metallic objects,
the real coins). Different coin types have disjoint version collections, and different coin type versions have disjoint collections of instances.
\item Between coin types and coin type versions there may be an intermediate level of coin type kinds
\item A coinage may be provided with a new coin type, or a coin type belonging to it may become obsolete.
\item Similarly a coin type may be provided with  a new version of that coin type and some older version may be declared obsolete. 
\item Banknote types, banknote type versions, bank not type version kinds, and banknotes (that is instances of banknote type 
versions and of banknote type version kinds) 
can be distinguished in a similar fashion as has just been done for coins.
\item With banknote family we denote the equivalent of a coinage for banknotes.
\item All instances of (subclasses of) Coin and Banknote are money-items. Money-items may be current or non-current (obsolete).
\item Informational monies add to this picture a variety of informational money items, often referred to as informational coins. Giving a classification of informational coins is an important issue lying outside the scope of this paper.
\item A moneyage combines a coherent coin family, banknote family, and informational money item type family.
\item A traditional money includes a moneayage, that is a coin family and a banknote family as well as a 
range of informational money-item types.\footnote{%
A money transcends over its underlying moneyage by the design of circulation processes (including bank regulation), 
minting methods and rights, and the role of
its governance bodies.}
\item In this paper we will not speak of coin classes, coin categories, coin sorts, coinage members, (coin family members), coin versions, although each of these phrases can be provided with a reasonable meaning as well. 
A similar restriction applies to banknote related terminology.
\item It is common practice 
that a money is replaced by another money in the same geographic area in such a way that simultaneously all instances of the whole
moneyage  become obsolete and an entirely new moneyage is introduced as a constituent of the new money.
\end{enumerate}

As an example: the Euro coinage contains the single Euro coin as a coin type, the German and French instances of this coin type
belong to different coin type kinds, each kind having a sequence of coin type kind versions as subclasses, and each coin type kind version
has a dynamic collection of instances. 

Missing from this account is that for instance the German kinds for all coin types and banknote types can be 
grouped together thereby obtaining a
coinage kind (given a coinage) and a banknote family kind (give a banknote family) both inheriting from a moneyage kind. 
Different approaches to the hierarchical 
structure of Eurozone coins and banknotes lead  to contemplating a matrix organization for which we could not find a 
convincing representation. An approach to this matter calls for the use of multiple inheritance, 
in spite of the neglect of that feature in traditional and popular object oriented program notations.

\subsubsection{Formal methods in money and finance (FMiM\&F)}
These techniques have been developed with an intended use concerning 
the analysis of digital machines, computer programs, and computing systems. When performing research of this style
in M\&F that work contributes to a subject area which we will term ``Formal Methods in Money and Finance" 
(FMiM\&F), which itself may be considered part of a wider area that may be 
referred to as ``Informaticology of Money and Finance" (IoM\&F). 
``Financial Logic" (i.e. reasoning about money and finance) 
will be a part of this field, and so is ``Financial Algebra" (calculations for finance and money). 
Both Financial Logic and Financial Algebra are also contained in the subarea FMiM\&F.

IoM\&F might comprise at least financial processes (in a formalized setting), 
financial data types (applied to financial statement descriptions), abstract money types (dynamic descriptions of forms of money
and modular specification of financial documents. IoM\&F also includes the theoretical part of computational finance.

\subsubsection{Brands of FMiM\&F and of IoM\&F}
The above description suggests a particular set of formal techniques to 
be of relevance to FMiM\&F and to IoM\&F. This, however, is a matter of taste.
Alternative bundles of technical (logical and mathematical) tools may also be employed in an effort to design a meaningful 
contribution to FMiM\&F and IoM\&F.  Moreover, the fact that these tools might 
simultaneously be applied to a coherent area (money and finance),
justifies a preliminary classification of FMiM\&F as a topic in theoretical informatics, and that automatically creates the 
subarea IoM\&F of theoretical informatics. We will work as if that 
classification of FMiM\&F is already in place, rather than to look for its  
classification as a subject in economy, in philosophy, in political science, 
or in sociology, each of which will probably provide far less fertile ground for deployment of said techniques.

Of particular importance might be the concept of loose specifications. 
In later section we will discuss virtual monies. Although formalization will not
be pursued one may imagine an axiomatic approach to financial systems, including its monetary component. 
These axioms may constitute
a loose specification admitting many models. Some models correspond to formalized ``normal''
financial systems with formalmoney, some models
may describe virtual financial systems containing virtual money. 
In a virtual financial system many aspects are skewed and the corresponding
formalmoney fails to satisfy the most common moneyness criteria. 
Besides these models there may be models for pseudo financial systems 
incorporating a near--money rather than a money. Such systems have a faint 
resemblance to financial systems only but the similarity with a normal financial system
is an illusion. 

\subsubsection{Formalism in economics}
In economics the label formalism is used to denote work that mainly consists of the mathematical analysis 
of a single and often quite simplified model.\footnote{%
That is quite different from the axiomatic style of  most formal methods in computing, 
which is often too general rather than too specific.}
The so-called post-autistic trend of thinking criticizes this particular styple of formalism as being pseudo-scientific 
(see Ardalan \cite{Ardalan2003} for a survey of paradigms in finance including post-autism, and \cite{Royston2004} for an
application oriented perspective on post-autism). Post-autism considers quite a bit
of mainstream economics unconvincing, mainly because of the perceived implausibility of various assumptions which 
underly equilibrium  theories.

Although the term post-autistic is quite unattractive the current paper might be classified under 
what Ardalan labels the structuralist 
paradigm which hosts post-autistic finance as well. However, what Ardalan refers to as the interpretive paradigm 
which also contains behavioral finance is
a more clearly demarcated paradigm in which our work on formalistic theory of money might also be fitted.

\subsection{Survey of the paper}

Here is a brief account of the successive topics covered in the paper.
\begin{itemize}
\item We provide a number of preliminary remarks on money and finance which set the stage for further developments. Some 
new jargon is introduced in particular concerning money classes and moneyages. 
These remarks assume a naive understanding
of money all of which will be put into question below.
\item In a similar style we provide a number of remarks concerning theories about money. Appreciation of there remarks
 requires that one accepts a naive 
understanding of the notion of a theory about money. 
\item Then we provide a meta-theory of definitions  in the form of a classification of definitions.
\item There is a formidable literature on M\&F in existence already. 
If one views the economic, social, legal, political, and philosophical investigation of money as a coherent and very significant 
running affair, the question can be raised in advance how to make a start in this area from a background of what might be 
considered merely a collection of logical and mathematical techniques taken from another area (computing)
without any proven specificity towards money and finance.

We will provide some suggestions for guidelines for performing research and writing papers in a field where one will unable 
oneself to take all relevant work already in 
existence into account, simply for reasons of lack of time. 
We will  assume that novelty and innovation are preferably perceived as virtues of work in 
progress or recently finished that may 
come about only on the long run rather than admitting definite assessment and judgement in real 
time. We will look into the following questions: 
\begin{enumerate}
\item Can one avoid making claims on novelty and priority which turn out to be wrong?
\item How to avoid ignoring relevant parts of preceding literature (also called prior art, 
in the jargon of intellectual property rights)? 
\item How to avoid ignoring relevant prior art as a 
consequence of working in a so-called community which among other functionalities may serve as an ad hoc coalition of 
authors who implicitly agree to focus on a limited part of prior art only?
\end{enumerate}
\item We provide a description of the literature on money and finance which may be considered part of theoretical informatics (including 
logic and discrete mathematics) already. A strategy for dealing with this body of literature when dealing with the research 
questions of IoM\&F posed before will be outlined.
\item We will provide a description of the existing theory of money and finance outside ToM\&F/IoM\&F (where / stands 
for minus). 
This description must provide some insight in how to go about novelty claims (both concerning questions posed and answers given) 
regarding ToM\&F/TIiM\&F. Included in this part is:
\begin{itemize}
\item A strengthening of the arguments given in \cite{BergstraL2013} for a dual money system 
that may implement interest prohibition more directly than is the case conventional Islamic finance.
\item An argument why Bitcoin may change the way in which mathematical theories of money approach the subject.
\end{itemize}
\item We will formulate a number of roles or identities each defining a perspective from which a theory of finance can be sought.
Rather than looking for the optimal theory of money for all or most current purposes 
we will assume that for each role a dedicated theory
of money is to be found which serves as an interface between individuals 
performing that role and the concept of money at large. In particular:
\begin{itemize}
\item We provide an informal theory of money for one specific role in particular: the SAR (subordinate administrative role).
\item We propose an interpretation of  ``virtual money'' and ``virtual near--money''
 such that it is plausible that a SAR specific theory of money gives rise to a special purpose virtual money or a virtual near--money.
\item We formulate a principle that expresses the tendency of top-management to have a local and corporate internal virtual near--money
deviate from a recognized money.
\end{itemize}

\end{itemize}

\subsection{Preliminary remarks on money}\label{SubsectPROM}
There is an amazing number of concise and compelling remarks reflecting on money available in ToM\&F. Some of these 
remarks are helpful in setting the stage for the task at hand. Here we collect a number of 
disparate statements some of which were found in the literature
that can help to locate with some precision the topic of money in a way that serves our further objectives. 

\begin{enumerate}
\item How to write about money if one of the conclusions will be that money is still in need of a comprehensive definition. Can
one say that \emph{money is sufficiently important to justify the effort to produce an up to date complete and consistent 
definition of it}?
If so, money has become quite important without having been properly defined, which reduces the urge to find a definition.
If not, there might be a circularity going on, as an attempt to provide a definition might subsequently 
reveal that money is more important than originally expected. 

\item To deal with this unclarity about the concept of money the notion of a 
`survey study definition' may be used. A  survey study definition of a concept X  provides a number of judgements by relevant
individuals of the form ``p belongs to X'' or ``p does not belong to X''. These judgements may be contradictory and then
 numbers count as if deciding membership of X were a matter  majority voting. Concerning money one may 
 collect a substantial number of votes for the assertion that the 1 Euro coin is an instance of money. A majority may vote against 
conceiving a credit card as a specimen of money in comparison to a singe Euro coin. From this hypothetical 
survey study we assume the following results:
\begin{itemize}
\item All interviewees agree that money exists, all count Eurocoins and banknotes from Eurozone member states as well 
as commercial bank demand deposits that allow free and immediate usage as money (independently of 
bearing interest).
\item All interviewees agree that money is very important and that it plays an unavoidable role in many processes.
\item Interviewees disagree on the status of cash cards, donation cards, debit cards, credit cards, cheques, foreign money,
government bonds, and savings accounts, but they all agree with the assertion that for each of these cases 
the classification as money is merely a matter of convention and that anyhow neither classification has any 
significant consequences. 
\item Interviewees don't know the phrase near--money but upon its explanation a significant majority of the interviewees
is comfortable with a classification of cash cards, debit cards, and foreign money as items of 
monies (that is forms of money) and of the other items as belonging to near--monies.
\end{itemize}
One may now conclude that money is worthy of further investigation. 
This amounts to the investigation of the class of objects and processes which people often 
refer to as a money or as a near--money. This is consistent with the potential, though
unlikely, conclusion of these investigations that ``after due investigation it has been concluded that money does not exist''.

This investigation may proceed extensively without before embarking on the construction of preliminary definitions of money.
\item \label{Diag} Probably no definition of money can both be valid and stable. 
As soon as a definition has been fixed, near--monies will be developed that explore and 
exceed the borders of the proposed definition. This is what we will call a diagonalization property of the concept of money.  
It shares that diagonalization property with the concept of a computer program notation.
\item There is no reason why a (specimen of a) commodity used as a money of exchange should have any other intrinsic value except being 
useful for the purpose of serving as a means of exchange, or as an intermediate (temporary) 
or long time storage of value. This was noticed by Simmel
\cite{Simmel1889} in 1889 as follows: 
	\begin{quote}Allein prinzipiell liegt nicht der geringste Grund vor, weshalb nicht ein 
	beliebiges Symbol f\"ur das Geld genau die
	gleichen Dienste als Wertmesser und Tauschmittel leisten soll wie Gold un Silber, sobald nur 
	die Uebertragung des Wertbewustseins auf jenes in 
	volkommenen Masse stattgefunden hat, was durch den Process der psychologischen 
	Emporhebung der Mittel zur W\"urde des Endzwecks sehr 
	wohl m\"oglich ist und auf anderen Gebieten hundertfach stattfindet.
	\end{quote}
\item Commodity money often exists in the form of coins. Fiat money\footnote{%
This contrast between commodity money and fiat money money can be criticized. 
Some consider banknotes and bank accounts to
be classified as commodity money just as well.} exists in the form of banknotes. Convertibility then 
means that the fiat money is guaranteed to be redeemable (by the issuing 
bank) for commodity money. Inconvertibility (of banknotes for coins)
is a sign of advanced development of a monetary system. Bank money (balances on demand deposits)
may be considered even more indirect than banknotes and equally
in need of convertibility. However, if one holds that moneyness is primarily related to liquidity then it is clear that all but the 
most simple transactions are best performed via bank money. Convertibility (for a form of money) nowadays means 
that assemblies of its instances are readily convertible to (very liquid) bank money at fixed rates of conversion 
(at least not involving some local market). However, if one enters a bank with some 100.000 Eurocent coins in order to 
settle a debt this plan may prove problematic. 

\item If one considers bank account money to possess the higher degree of liquidity in comparison to ordinary currency 
(coins, banknotes, electronic cards for payment, debit cards), then some further aspects need attention.
\begin{itemize}
\item Under this assumption the best money to which the public has access is all 
managed by the commercial banks and therefore the risk of bank failure needs to be taken into account. 
\item That makes bank accounts with the economically
strongest banks the best available money, at least if the money is to be used as a store of value. 
\item This state of affairs makes it imperative that the pubic can assess the relative strength of 
banks. Even if a state guarantees an account (at least up to some amount) in case of bankruptcy of a bank, 
because of potential delays in redemption and also because of
potential legal problems when claiming from the state, a bank failure in general poses a liquidity risk for its account holders.
\item Each bank may impose upper limits on the size of its accounts. This policy induces upper bounds on homogeneous 
(that is being assemblies of the same money class) quantities of money which may exist.
\end{itemize} 

\item \label{CCneeded} Bank money is at the top of the hierarchy for large payments only (say over 10.000 Euro). For
smaller (though not very small)
payments and for purchases  where the seller is confronted with a risk that cannot easily be 
insured during or after  the transaction credit cards are more liquid. For some transactions such as renting a car
at an airport the use of
a credit card may be necessary and no ordinary form of money may be accepted. One may wonder 
whether such cars are indeed for hire. Clearly there is more to credit cards than merely providing a 
service on top of a bundle of existing financial products. 

Credit cards are related to creditworthiness. Having a fixed income is helpful if not needed to acquire a credit card.
Some transactions (for
instance renting an apartment in Amsterdam) require a proof that the prospective inhabitant enjoys a  monthly private 
income, and that proof cannot be replaced by any other form of guarantee. Again one may 
question whether such apartments are actually for hire.\footnote{%
Two years later when revising this paper with the crisis on the Amsterdam housing market being felt in all market segments, 
landlords have become far less demanding.}

\item The term banknote refers to times when an individual commercial bank had the right to `mint' its own paper money. Today
banknotes are in the majority of cases issued by a national bank or  a central bank which exercises a monopoly 
for that purpose. The term banknote is anachronistic (just like ``digital coin'' is anachronistic), it should rather be
``national bank note'' instead. Banks issue bank accounts on their own terms
in the same way as they used to issue their own banknotes. Because bank money has become so important it is nowadays an
impractical vulnerability that bank accounts are not simply ``national bank accounts''. 
Some steps are taken in that direction with governments forcing commercial banks into a 
collective insurance scheme which protects bank account holders (up to some 
limits) against the bankruptcy of the bank where they hold an account.

\item Outdated forms of money should also be considered money. Thus money is what has been money, 
as well as what is money, and what might have been money. 

\item Cowrie Shells constitute a money class, a subclass of commodity moneys, 
though that class is obviously not a part of the family of money classes
that are used in the Eurozone around 2010. If one considers gold bars of a specific weight specimen of a money class during 
some episode (which is
historically not too far-fetched) then one should also acknowledge that in a later episode the 
same bar is not money anymore but it has 
now become merely a valuable commodity instead. 
This has happened with many coins that have been specimen of live coin types 
and thereafter have become valuable
commodities just because of previously having played that role. 
We may speak of Eurozone 2000-2010 money classes (or of Eurozone 2000-today 
money classes), and then the gold bar is not a 
member of any of those. 

\item Any money class may be called valid (at some moment and place) if it can be used as such and invalid if it is either 
outdated or perhaps not yet 
valid while still in phase of design and development. In any economy the family of valid money classes evolves in such a 
way that classes are removed and that new classes added.
As a part of this evolution but of a secondary importance subclasses within classes can be retracted or introduced.

\item Money classes can be categorized in meta classes (categories). 
Coinage may be considered a  money category, just as banknote families. 
As a category Coinage is not (yet) outdated, but most instances of Coinage (coinages) have become invalid
and so have most individual coins (instances of coin types). Harris
explains in \cite{Harris2006} that Romans knew banks and bank accounts as well as (non-local) 
bank account transfers for the purpose of trade. 
Consequently current bank accounts are instances of a money category with a long history. 
Like with coins most historic forms of bank accounts have lost their validity.

\item The evolution of money has a layered structure with three levels: 
instances (specimen), money classes, money categories. 
Taking a closer look one notices that:

\begin{itemize}
\item The process of replacement of categories is very slow. 
We may just observe the historic fact of the phasing out (after 2500 years) of metal coins in the near future. 
Digital money-items arrive 
as a replacement, constituting a different money-item type with a comparable functionality. 

\item The process of money class replacement within a given category take place at high speed: 
the life time of a design of coins and banknotes (i.e. the overall design of a Currency family) seems to 
be less than 50 years.
\item Money class replacement has a fine structure: 
the entire system may be redesigned providing new money classes for all categories at the same time (the
introduction of the Euro is an example). 
More frequently individual money classes are redesigned or introduced. (The introduction of a 1000 Euro banknote
need not imply a major change of any other money class of the Eurozone moneyage.)
\item The replacement process of instances (specimen) of individual money types is faster once more. 
Banknotes do not easily survive 25 years of circulation.
\end{itemize}

\item Having available the term moneyage, as introduced above, as well as the viewpoint that bank accounts may constitute 
informational money 
types in a moneyage one may ask whether interest is a necessary feature of some of the moneyage types. 
We will deal with Islamic money below in 
more detail, but clearly when full generality is aimed at, interest cannot be a required feature of a moneyage. 
However, it makes perfect 
sense to imagine a moneyage in which some money classes generate interests as a required feature in order 
to accommodate a particular rendering of non-Islamic finance, 
whereas a subfamily of moneyage types, none of which generates (or requires) interest payments, is also present in order to accommodate
an adequate (sub)system for Islamic finance.\footnote{%
If interest generating liabilities are not counted as money 
the relation between money and interest is indirect
and the Islamic theory of money can be understood as being less remote from its non-Islamic counterparts 
than usually thought.}
\item \label{Cmcf} An amount of money always exists in a current moneyage. 
An amount can be imagined for instance as having been prepared just before 
performing a transaction (that is preparing the amount as a means of exchange),
or as having been observed implicitly and retrospectively for the purpose of taxation. 
It can also be imagined as an amount received after
an exchange of goods, commodities, services, or of promises to provide either of those in the future 
have been delivered to another party. 
A moneyage provides a family of moneyage types, which we will refer to as the current moneyage type family (CMTF). 

Any amount in terms of a given CMTF is composed of parts each of which belong to the same 
currency type or informational money type from the CMTF at hand. 
Each part may consist of one or more objects.
A bank account is often a single object, whereas a quantity of currency (in the form of coins and banknotes)  
often appears as a plurality of objects of the same class.
\item \label{Amount1}More often than not, an amount prepared or received cannot be 
observed directly in the sense that a larger amount is in fact prepared and 
only by actually performing the exchange the required amount is singled out. 
After an exchange the receiving party may also have an indirect perspective 
in the sense that it is the increase in value of amounts before and after the transaction that matters. 
Due to the heterogeneity of amounts,
caused by the variation of currency types in the CMTF, 
it becomes imperative to use a money of account in addition to a money of exchange just to
calculate the value of amounts.
\item \label{Amount2} Assuming that in all practical cases a moneyage may have a CMTF 
consisting of at least 10 (and often many more) moneyage types the money of
account functionality becomes necessary precondition for the money of exchange function 
because to determine the value of an amount a computation is needed. 
More likely than not, in all but trivial
 cases that computation cannot be done by head. 
 This leads to a slightly more detailed definition of an amount: it consists of  a triple of a unit (say Euro), a value,
say 15.350.21 (the last two digits representing a percentage of a Euro), 
and an amount in the sense of the above item \ref{Cmcf}.
\item In practice both definitions in item \ref{Amount1} and in item \ref{Amount2} are defective. 
Although an amount received seems to be a stock quantity
if it is to be found on a bank account it is usually observed via corresponding flow variables. 
That means that by indicating the logging of an incoming 
transfer and the quantity it carries one observes an amount received. 
Comparing successive values of a single account may be problematic if
many transactions with different parties or concerning different exchanges with the same party take place concurrently. 
Thus an amount prepared before transfer, an amount transferred, 
an amount received, and an amount retrospectively observed in the past,
each have subtly different technical definitions. Most of these notions or mechanisms
refer to processes or progressions of processes rather than to static data or static 
configurations of amounts in the sense of item \ref{Cmcf}.
\item From the perspective of formal methods in computer science these differences between distinct shapes of the 
seemingly obvious notion of
an amount are quite significant. Each error in a precise formulation of one of these concepts
may cause deadlocks or even quantitative mistakes to occur in an automated system that is supposed to support 
the implementation of storage and exchange of amounts of money.

\end{enumerate}

\subsection{Preliminary remarks on theories of money}
Because so much has been written about money, one can hardly write about money in a reliable way without writing about 
theories of money. Ideally one should first become fluent in the major theories of money and only then try to write,
if at all. However, this seems to be unfeasible in the light of the size of the existing literature about money. 
We will return to that  hindrance in a later Section. We now list some observations about theories of money first.
\begin{enumerate}
\item Theories of money are almost 2500 years of age. Different viewpoints have always existed. The 
ramifications of viewpoints on money since say 1935 are formidable.  Fontana writes in \cite{Fontana2000} that `No one 
aspect can stand on its own as a complete account of what money is and what money does in a modern economy'.
\item Theories of money concentrate on areas such as: private investment, business management, creation of money by banks,
taxation, government
spending, creation of money by the state, role of central banks, cooperation and competition between commercial banks, 
international  cooperation between
national banks within currency areas, international financial systems and structures. 
\item A dominant dichotomy in arguments consists of ex ante versus ex post. Economic science performs at its best 
with ex post explanations
(that is explanations of past events.) Ex ante arguments, which should produce predictions with a meaningful degree of 
reliability are much harder to
develop. It is unclear to what degree the theory of money supports any ex ante reasoning at all.

Hicks \cite{Hicks1962} states that prediction is not what economists (or financial theorists) must 
attempt to do. Instead economist should explain concepts, in the cited paper
he takes liquidity as an example of a concept that may be frequently 
used in a political context and still is in need of clarification.

\item In recent centuries  financial technology has been innovating quite steadily. Ex post explanations often have to take 
into account structural modifications in the financial system. 
Because ex post conclusions are most susceptible to validation and with
recent history (relative to the moment of performing research) invariably demonstrating innovations or structural changes,
 historical analysis has become a major tool of investigation in the theory of money in the last 200 years.
Econometric data collection about fairly recent processes and events may be considered to constitute merely an
extreme case of historic analysis. 

Mitchell \cite{Mitchell1896} provides an example of these effects: 
writing in 1896 he criticizes Ricardo's quantity theory (then almost a century of age) by means of conceptual analysis 
and by providing historic data from 1861 onwards. For Mitchell, bank accounts are definitely not money. He acknowledges 
the existence in the USA of those years of nine forms of money, and several more means of exchange. There is a very 
precise count of (then) base money available. 
Mitchell needs to formulate and defend his critical position towards Ricardo's quantity theory while
in Ricardo's time bank accounts were far less important and with the understanding that 
Ricardo made his point under rather strict
assumptions about state control of money (no free minting for instance), which were not valid in the USA when 
Mitchell wrote. Mitchell's intellectual 
opponents, on the other hand struggled with an outdated version of quantity theory in a changing world. 

\item Many works on money combine an analysis of money with
an appreciation of its assumed relevance for or application to major political or social problems.
Remarkably many papers seem to put forward that either the cause of or the solution to the problems of mass unemployment
are to be found in an adequate understanding of money. 
Lerner \cite{Lerner1947} is an example of this style of writing, by putting an emphasis
on the control over employment which a state obtains by having control over money, and at the same time discussing the 
principal origins  of money. The lack of housing capacity or problems with healthcare provision are also
but less frequently connected with theories about finance. There is no proof, 
however, that an ex ante understanding of money (as developed
by inspection and comprehension of scientific literature on money) is a decisive
factor in solving such grand social problems. It may be a very useful factor for avoiding known mistakes, but in new 
circumstances the conflicting imperatives from different strands of the theory of money hold no more authority 
than the persons advocating these different strands hold themselves. 
In \cite{Vickrey1998} on finds a critique of a number of `fallacious' arguments from financial theory to policy.
\item Money plays a role in the international competition between states and between
coalitions of states. There seems to be no ground for the assumption that the theory of money provides decisive ex ante 
arguments about the validity and the perspective of competing policy proposals. 
The financial system is about the organization of relatively unconstrained spaces of 
`free' decision in the presence of uncertainty (that implies the inability to analyze future events faithfully 
in probabilistic terms) for a number of different parties involved. It seems to be 
the case that the following phenomena, which seem at first sight  to represent weaknesses of a financial system,  
all add to the value of money both as a means of exchange and as a store of value:
	\begin{itemize}
	\item The existence of different future policies
	(possible and contemplated) about money. The very presence of such significant degrees of freedom itself adds 
	to the value of money per se (just like
	stock market volatility increases the value of stock options). 
	Financial derivatives provide an expression of that form of value.
	\item The intrinsic difficulty of assessing what constitutes money, 
	and the consequential difficulty of assessing the quantity of it resulting in the
	fundamental difficulty to validate or refute up to date forms of quantity theory.
	\item The difficulty, if not impossibility to predict future rates of inflation.
	\end{itemize}
\item Yeager writes in \cite{Yeager1985} 
\begin{quote}Figuring out ex post how money should have been defined and regulated is not the same 
as knowing how to do  so currently.
\end{quote}
The difficulty was obvious to Andrew in \cite{Andrew1899} already. Andrew emphasizes
the point that `according to the popular opinion' one accepts as money those 
\begin{quote} ...media of exchange which circulate without the 
necessity
of indorsement, or of registration in books, or of conformity to any other condition than the mere transfer of the certificates of 
value from one person to another.
\end{quote} This focus on the autonomy or owner independence
 of money-things has become quite rare in the literature on money. It
seems to have been abandoned. Kepner \cite{Kepner1959} discusses in amazing detail the 
development of joint bank accounts each version of which may 
need its own assessment concerning its moneyness. 
Tobin \cite{Tobin1961} indicates how well-known theories of money are dependent on
a selective view of which bank accounts (closed money substitutes of whatever maturity) are considered money.

These complications exemplify the diagonalization property of money mentioned in remark 
\ref{Diag} of the preliminary remarks on money. It is plausible that obtaining a watertight current understanding 
of financial policies is equally unachievable. An ex ante
prediction of such policies is a matter of pure speculation, and so is the ex ante specification and demarcation of what is money,
irrespective of one's favorite theory of money.

\item Although in comparison with coins, banknotes, and bank money, credit cards are new and somehow 
exotic, credit cards require our  attention. 
Outstanding debt
with the credit card industry is huge but hard to measure (see \cite{Zinman2009} for the USA situation where total 
credit card private debts
are claimed to be of the same order of magnitude as national debt per 2006).  In \cite{GansKing2002} the Australian 
credit card system is analyzed in view
of imminent system changes. That work is based on the assumption that 
\begin{quote}.. it is the customer who determines the choice of payment 
instrument for any specific transaction; a choice that may impact upon the 
payoffs and profits of other participants to a payment system.
\end{quote}
Because:
\begin{quote}
As we will demonstrate, this insight is sufficient to both give some weight to concerns about 
inefficiencies in credit card associations but also to isolate the key impacts of the policies that will be enacted in Australia.
\end{quote}
However, as we mentioned already in item no. \ref{CCneeded} of the preliminary remarks on money,
credit cards may be more liquid than both currency and bank money (with car rental as an example). Not only
are credit cards liquid because of a credit line established by an independent party (the issuer), 
but more importantly this issuer can guarantee 
that subsequent payments will enabled when needed (at least for important customers such as car rentals and hotel chains). 

Kahn and Roberds  \cite{KahnRoberds2002} discuss finality as an important criterion for the 
effectiveness of payment transactions. Finality is comparable to atomicity of database transactions in computing. 
An atomic transaction is either performed  till completion or not at all. Credit card
payments provide a very high level of finality to payees, better than currency (banknotes might be false, 
coins may have been stolen
or counterfeited), cheques (which may not be redeemable after all), 
debit cards (no guarantee that directly related subsequent payments 
can be performed.\footnote{For instance the minibar in a hotel, or fuel for a rented car.}) 

Ellmann \cite{Ellmann2009} asserts that the phenomenal growth of the credit card industry has forced significant changes 
in USA bankruptcy laws with as a consequence that in his words:
\begin{quote}Indeed, for many Americans credit is now income.
\end{quote}

If one considers the entire credit line available to a credit card holder a constituent of his liquidity, 
the credit card system increases 
total liquidity by the sum of volumes of unused credit lines for all card holders and all card issuers. 

If credit is more liquid than money, the whole system of concepts seems to become circular to the point that 
comprehensible definitions may be hard to produce. For instance now a higher liquidity preference may induce the intention
to use money instead of holding it with the objective of improving one's creditworthiness. 
From a computer science perspective this form of recursion may be attractive, however, because it may be considered 
a challenge for the application of semantic techniques.
\item Many foundational papers about money take significant positions regarding the history of money. 
An adequate appreciation of the history of money seems to be very hard to acquire. Wray, probably the most prolific author on money since 1960, with (\cite{Wray1993}) provides an example of a foundational
paper which sells some historic assessment: rather than the orthodox view that commodity money 
(usually seen as exogenous money) historically 
gave rise to endogenous money creation (credit money as nowadays created by commercial banks)
according to Wray it has been just the other way around.

\item Also for papers aimed at an explanation of current phenomena the display of a historical survey is not uncommon. 
For instance in
\cite{DwyerSamartin2009} the question is posed why banks promise to offer paying par on demand for certain liabilities while
fractional reserve banking will cause them to be unable to live up to this promise sooner or later. The issue turns out to 
be so complex that a historic perspective is helpful for structuring the matter.

\item A common strand of theory about money puts the state at center stage and considers the value of 
money to generated and maintained by the state's stated intention to allow taxes being paid (redeemed) via said money. 
This analysis is often labeled the state view of money.
Given the potential relevance of taxation for the very existence (and according to some authors 
even the historical coming about) of money it is remarkable how little technical information about taxation
is taken into account or used in foundational papers on money. For instance VAT is absent from the principled writing on money. 
The hypothesis
that money is helpful (if not necessary) for effecting taxation even in a barter economy seems to play no role in theories of
money. Nevertheless the fact that money allows a uniform approach to taxation on individual transactions (whether involving 
barter or not)
seems to be of paramount importance for the state's interest in financial technology.

\end{enumerate}
\newpage

\section{A classification scheme of (imaginative) definitions}\label{Definitionclasses}
Irrespective of the extensive amount of previous work on money we 
consider the design of definitions for monies and moneyages  as a research problem which still merits further attention.

In order to deal with those maters we will develop a theory of definitions which applies to the case of monies. In
particular we will ``coin'' the concept of an  {\em imaginative definition}, and we claim that imaginative definitions
of monies are worth of being developed. 

Theories
of definitions belong to philosophy of science and ought to be take from there in principle. We failed to find an account of how
to give definitions that simplifies the task in the case of money and we have taken that failure as an incentive to produce a home made
design of a theory of definitions. Before embarking on the meta-definition of definition in Section \ref{Imdefs} below we list 
an number of issues concerning money, al of which depend on having appropriate definitions at hand.

We notice that M\"aki (\cite{Maki2004}) provides a rather philosophical but nevertheless 
explicit meta-theory of definitions in support of a 
definition of money. As an application of his point of view M\"aki discusses the important 
question whether or not an entire community can 
be mistaken in the judgement that they
are making use of money.\footnote{In \cite{BergstraL2013} the case is made that the Euro might be 
considered a near--money rather than
a money in a dual (near--)money design including besides the Euro a modified version of Bitcoin called Bitguilder. The rationale
for the dual system being that it avoids interest payment on money, while admitting interest payment on near-monies.}

\subsection{Is there a remaining definitional problem for money?}
We mention a number of aspects of money and of its use that may call for a more systematic 
approach to definition from the point of view of semantics in computer science.
\begin{description}
\item {\bf{circulation of money.}} One needs to define in what sense there is 
circulation of money, what kind of topology must be assumed. 
Is that notion still applicable with digital money? How to deal with credit money? 
Circulation theory provides the underpinning of conservation laws
for money. 
\item {\bf{circulation velocity.}}
It is illuminating to compare the notion of (money) circulation velocity  with the concept of execution speed for 
computer programs on computers. Now a meaningful definition
of execution speed for a program requires a thorough definition of the 
execution of a program at the first place and such definitions are not easy to 
provide. Once the necessary details are added intuition gradually degrades. 
In the case of money circulation velocity of coins is a different matter from 
circulation speed of bank accounts.  For any application in economic theory
one needs to define an average circulation velocity for all simultaneously existing instances 
of each money class during some time interval. For coins and banknotes the 
phenomenon of change may lead to irrelevant movements that 
induce
an overestimation of the `intended notion of circulation velocity'. This latter definitional task
is of comparable difficulty to the problem of defining execution 
speed for a computer, say a multiple pipe-lined machine architecture running a multi-thread. 
Such definitions have turned out to be
notoriously hard to provide (see e.g. \cite{BergstraMiddelburg2007b}).

In any case circulation velocity is an endogenous, i.e. a property that can only indirectly be influenced property by modifying
exogenous properties according to many authors.
\item{\bf{physical location of money.}} The concept of physical location of money class instances 
is not even easy for coins and banknotes. It becomes
more difficult with bank accounts and electronic monies. 

\item{\bf{Formaleuros and formalbitcoins.}} If one intends to develop theory about a particular money such
as the Euro or Bitcoin, then one is interested in stating facts which are more general only to specialize to
actual monies in a final stage. Most reasoning is performed about a formal model of a money (and its money-items) 
rather than about a money proper.

Logical or mathematical definitions of say Euro or Bitcoin, don't produce Euro or Bitcoin proper but rather formalizations
of these monies at best, say formaleuro with unit FEUR and formalbitcoin with unit FBTC.\footnote{%
FEUR and FBTC can be used as units of account in models of management accounting. In a model of physical coin 
circulation involving 1 EUR coins, one will speak of the circulation of 1 FEUR coins instead. As an advantage of this cautious
and indirect terminology we mention that questions about wear of a specific 1 FEUR coin make no sense unless the feature
of wear has been incorporated explicitly in the definition of formaleuros at hand. 

In the context of Bitcoin speaking of FBTC transfer with quantity $q$ from an account $a$ to an 
account $b$ can be done without any risk that this utterance is 
interpreted literally (as referring to a BTC transfer that has taken place of may take place in the future), 
even if $a$ and $b$ constitute a public-secret key pair that has been used or 
might still be used in actual Bitcoin practice. Remarkably, by publishing a key pair $(a,b)$ in a formal paper that has been 
phrased in terms of FBTC
only, that key pair is compromised upon publication the future for a user (say $P$) of 
Bitcoin who must always be on the lookout for intruders who check 
$P$'s public keys against all conceivable secret keys they have ever seen, and that collection may then contain $b$.}
Such formalizations are both more abstract (under-specification) 
on certain aspects, in particular about all aspects related
to physical presence of entities, and more specific (over-specification) when dealing with mechanisms (in particular when
physical interaction must be modeled with logical means).

Thus: defining monies as pieces of mathematical and logical work, even when performed at an informal level, cannot
go beyond the production of formal counterparts of monies. About the formal counterparts one may reason and 
the translation of patterns of reasoning to the ``real case'' needs to be performed with great care in each occasion.
\item{\bf{clarification of formalcoins and formalbanknotes.}}
Assuming that coins and banknotes are analyzed in formal terms, by way of instructive  so-called 
imaginative definitions, that process gives rise to formalcoins and formalbanknotes. 

The following general questions can be posed for each formalized coinage
and depend on sharp definitions that clarify how and what of these particular money classes. In fact the quality of 
imaginative definitions can be measured to some extent by the way in which these 
questions can be provided with reasonable answers:
\begin{enumerate}
\item Assuming that coins are coming to an end in the electronic age the question arises: 
which aspects of the service provided by coins can
be formulated abstractly and in such a way that they may survive a drastic technology 
change. What is a technology independent formulation
of the virtues of coins (and banknotes). 

Perhaps this question is silly. Comparing means of exchange with means of transportation: 
there seems to be no point in carving out the precise combination of services 
provided by riding a horse. Modern 
transportation theory has decomposed that package  new packages around 
transportation (bicycle, motorized car, hot air balloon, and so on, have been developed instead.\footnote{%
The following aspects may enter this kind of discussion: coins have a form of independence and stability that is lacking 
for other forms of money. At the same time they are very vulnerable to theft. Coins 
are transported by an owner and offered on site of a transaction. Coins can be stored and do not degrade. 
Coins are unharmed by water and by most other forms of contamination. The size, weight, and shape of coins
can be psychologically satisfying.})

\item When is a formalcoin `false'? This is far from obvious. In fact the definition of a formalcoin needs to be expanded 
with a sufficient number of 
attributes for providing additional structure which permits any statement about fraud or forgery.

\item Once formalcoins and formalbanknotes have been specified:  is there a clear logical distinction between the two? (That 
seems to be the case: assuming that a series number is printed on each formalbanknote then given
two identically looking formalbanknotes at least one the two must be forged. This inference rule has no counterpart in the 
case of coins as these do not carry series numbers.)

\item When multiple moneyages  are considered we prefer to write FEUR instead of formaleuro, FUSD instead
of formaldollar and so on. 
What is the type structure of money. Are say FEUR and FBTC  instances of 
a wider class formalmoney. Are different forms of money, say 
credits and debts also to be classified in this type structure? How do different national forms of Euro 
formalcoins fit in this picture?

\item Can a coin (or rather a currency item) be defined in such a way that it consists of information only. 
If so is it conceivable to have a theory of money in which money only features as information.

\item How to specify collections (hoards) of coins and banknotes such as occur within a wallet. 
How to specify methods and algorithms for search and retrieval as a part of payment processes. 

\end{enumerate}

\item{\bf{identification of abstract moneytypes.}}
Formal money classes may be alternatively called abstract money types. 
In practice abstract money types are used when designing so-called financial products.
Developing abstract money types for  bank accounts is a non-trivial matter. Indeed abstract money types are quite complex in 
comparison to the abstract data types that the computer science literature has on offer. 

Many questions can 
be put forward once one has decided to explain a bank account (below also referred to as product)  in formal terms.
Here is a listing of issues that come into play when an abstract money type that captures the essence of a 
deposit bank account is to be defined.

These issues per se are not research questions of course, rather their existence and number indicates that 
formal specification techniques from computing may be helpful to achieve the required levels of 
precision.\footnote{The question which (abstract) bank accounts should be counted as money is far from trivial. As it
turns out this changes in time, the collection of money classes grows at cost of the collection of near--money classes. Further
there are significant differences between Euro, USD and so on.}
\begin{enumerate}
	\item should there be an upper bound to the size of the amount and to its length of existence, 
	\item can the account have a negative value,
	\item what is the role of time, 
	\item how to specify an interest mechanism, if any applies, and where is interest transferred to,
	\item how to deal with restrictions on withdrawals, 
	\item are automatic withdrawal mechanisms a feature of the product or just of its use,
	\item are insurance policies against bank failures part of the product, how to deal with 
	erroneous or fraudulent withdrawals and additions, 
	\item what happens if the hosting bank leaves business, splits, or merges with another bank,
	\item is it important that the host is a bank, or can an organization different from a bank also provide bank accounts. If it
	is essential that the host of the account is a bank, what does that mean or is none of such 
	information a plausible part of the definition of an account,
	\item how to deal with information about transfers, 
	\item are periodic surveys part of the product, (if so what relevance has the form of delivery),
	\item should there be an online information system about it, if so is that password protected and what form of authentication
	will bee needed,
	\item is the product in essence independent of the ways in which it can be used,
	\item is the product logically dependent on any its predecessors (for instance a naming history of the hosting bank,
	number history of the account number),
	\item which authentication policies constitute a part of the product,
	\item can the product continue to exist during and after a change of identity of its owner, and if so, which part of the change
	protocol is viewed as functionality of the product,
	\item is an owner history part of the product,
	\item does the account still exist after is has been closed, and if so, how long,
	\item can there be more owners, or several persons having different rights of access and use, 
	\item is the complete history part of the product in an abstract sense even if the information is not 
	preserved by the provider of the product,
	\item given this large variation in options, can a core be distinguished which needs to be in place so that further aspects
	can be added to a variable degree. Is such a core amenable to a mathematical definition (formalbankaccount).
\end{enumerate}

A formal approach to money types may be helpful to distinguish money classes from neighboring near--monies. 
Indeed whether or not a specific type of bank account is to be  considered money may be dependent on its 
various structural parameters in non-trivial ways.

\item{\bf{security issues for financial products.}} For each abstract money type  
appropriate concepts relation  to security, anonymity, and privacy must be defined. 
Perhaps security must be even elevated to the most important level of the 
definitional efforts for formal monies.
\end{description}

\subsection{Imaginative definitions}
\label{Imdefs}
An imaginative definition provides a theory which can  produce a mental picture of a subject area in advance of any
confrontation with practice. An imaginative definition produces an imagination of a notion or a type of 
artifact independent of the associative
connotations that emerge from a person's experience with examples and instantiations of that notion. 
Imaginative definitions are not models of a reality
already imagined. Instead an imaginative definition provides an imagination 
(mental picture, conceptual model) which can serve as a point of departure for
subsequently dealing in a critical fashion with real phenomena. 

Appreciation of the importance of imaginative definitions is a matter of taste. We consider the development of 
imaginative definitions to be an important issue
for a range of themes. Imaginative definitions are designed on the basis of a loose set of intuitions 
concerning a family of related concepts where one
concept (or perhaps a few concepts) has (have) been singled out as the target for providing an imaginative definition. 

Each formal model of a theme containing instances of some concept automatically constitutes 
a candidate for providing an imaginative definition of that concept. 
Imaginative definitions for some concept can be developed in successive stages. 
A major reason for rejecting or for intending to improve upon a 
certain (candidate) imaginative definition is that it either lacks essential or characteristic 
information or constraints (example: a computer program is a bit sequence) or that it is too specific 
(e.g. a computer program is a Java text). Overly specific imaginative definitions can be so in 
different ways: a concept may be quite heterogeneous and some of its strands may have 
been captured while other strands have been entirely missed 
(example: a game is an interactive computer program such that ..., 
thus missing out on lawn tennis episodes and chess and many other meanings of game), or alternatively
it may be the case that a clear image is cluttered by unnecessary detail (example:
a computer program is a sequence of ASCII characters; 
even if one agrees that programs are to be defined as bit sequences insisting that these are
made up from ASCII characters may be considered too specific).

An exemplary example of an imaginative definition is the mathematical definition of three dimensional Euclidian space. 
Another imaginative definition is that of the real numbers and the continuum. 
These definitions are not serving as a 
model of something else. Rather such definitions help with creating a conceptual scheme on the basis of which further 
contemplation of models can be performed.

\subsection{A classification scheme for definitions}
With the objective in mind that one intends to develop imaginative definitions we will develop a classification scheme for definitions
based on experience in TCS.

Defining notions and concepts within theoretical computer science as within any other field of research is 
problematic in the sense that it may not be 
clear in advance what is to be achieved by providing a definition. 
Many different definitions for the same concept may exist and many grounds may be put 
forward for being satisfied or for being dissatisfied with a particular definition. Indeed
there may be as many definitions of a computer program
as there are programmers or programming teams around. 
Or perhaps only as many definitions as there are authors of books on computer programming, or merely as many 
as there are authors of research papers on the theory of computer programming.

Apart from this multitude emerging from the plurality of users of a concept there is also a 
divergence in objectives and criteria that may bring about a multitude of possible definitions. 
Writing this classification of definitions  has been triggered by an attempt to find a definition of the 
well-known phrase operating system (OS).
According to \cite{Middelburg2010} there is no precise definition of that concept available 
within the existing computer science literature. 
This is quite remarkable because is seems to be indisputable that operating systems 
are among the major deliverables of the computer 
industry and because the phase operating system is so commonplace in the computing literature.

\begin{description}
\item{IDBR:} {\em Informal descriptions by role.} A concept X may be defined by means of a description of the circumstances 
where concept instances (X's) play a role. The description can be provided
by means of informal explanations of that role as well as of criteria to be satisfied by 
instances that meet the requirements of the role at hand, 
of objectives to be met and of variations thereof. Methods of production for X's (factories in computing terms) 
can be taken into account as well as quality 
measures and accounts of historical development and evolution. All concepts of practical relevance admit this kind of definition.

\item{LSCD:} {\em Logical solitary concept definition.} An X is defined as the element of a logical/mathematical 
class (set, collection, category) of X's, the type of X. The definition comprises sufficient as well as necessary criteria for membership. 

Additional remarks:
\begin{itemize}
\item If the criteria are merely necessary then the definition captures a larger class and that should be given a different 
name. 

Example: consider the candidate definition: a program is a nonempty ASCII character string which contains some non-space symbols. 
Even if one admits that all programs are ASCII character strings, which is a conceivable though not a compelling
point of view, it should be admitted 
that such is not a sufficient condition and in fact a larger set is being defined.  The condition may be necessary but it is 
insufficient. Instead of speaking of an LSCD providing necessary conditions we 
may as well speak of a super-LSCD. A super-LSCD determines a super class of the intended concept. 
(That is a class having more instances than the intended concept.)
\item If the conditions are merely sufficient it is likely that a sub-concept is being defined and again a more refined naming is 
needed, e.g. P X's are .... This leads to a sub-LSCD (that is an LSCD of a subconcept, i.e. a concept with potentially fewer instances).
\item If the concept has first been given an informal description by role (IDBR)  then a proposed 
LSCD definition may be considered unrealistic just because it fails to capture the physical reality 
of the concept instances at hand. This is lack of precision is captured by the prefix L of LSCD. 

The LSCD can be used to reason about objects in reality for which in logical and mathematical terms LSCD provides a useful model. 
The usefulness of an LSCD in a particular setting cannot be guaranteed (in principle) 
on the basis of internal properties of the LSCD alone. 
Again considering the example of computer programs, suppose that an LSCD (say DefP) of ``computer program'' has been found. 
If P satisfies DefP then P is a mathematical entity. That implies that it can never impact the behavior of some physical machine. 
At the same time it is quite possible that as a part of an IDBR one has claimed that programs are intended to 
control the behavior of computers in useful ways and so on. By insisting that programs are mathematical objects one invokes a 
counterpart to the so-called body mind problem for the dead machines to be controlled by a program.
\item A particular complication is that LSCD's of the same concept (in terms of its informal description by role) can have 
different levels of abstraction. As logical concepts these levels of abstraction should be explicitly distinguished and provided 
with dedicated names. 

\item {\em Example.} The Nakamoto architecture as proposed in \cite{BergstraL2013} constitutes an informal version of an LSCD
for a class of Bitcoin-like monies. It is easy to transform that description into a completely formalized text which might serve as a 
formal definition of a particular class of peer-to-peer informational monies which includes Bitcoin.

A significant complication that was encountered when developing the description of the Nakamoto architecture 
is that resulting description not only constitutes an (intended) abstraction of Bitcoin but
at the same time it embodies an over-specification of its mechanics. The objective to capture Bitcoin mechanics at a particular level of 
abstraction seems to bring with it an incentive to specify simpler mechanisms that are supposed to bring about 
essentially the same functionality while working technically in a somewhat different way.
\end{itemize}

\item{SCFD:} {\em Stratified concept family definition.} In an SCFD a concept is positioned in between several other concepts. 
A stratification indicates that some concepts in this family are more central than others. The central concept (or concepts) 
needs (need) to be provided with an LSCD, each provided independently, whereas less central concepts can be defined (specified) 
by means of super-LSCD's (in order to make the definition more robust) and in some cases even by means 
of sub-LSCD's (usually in order to make the definition simpler, while leaving the generalizations to less strict 
sub-LSCD's as a task for a later occasion).

Example and remarks:

\begin{itemize}
\item As an example we consider the execution architecture (local environment of use) for a program. 
An SCFD for a computer program might in addition to an LSCD for ``program'' add a super-LSCD for a machine that 
may run the program, a super-LSCD for a local environment in which the program is used, and a sub-LSCD for a 
network in which this machine is operating as well as a sub-LSCD for a program library from which programs are 
taken and a sub-LSCD for a configuration management system in control of that library. Further there may be a 
sub-LSCD for a run of the machine (as a function of behavior of the environment, and admitting infinite runs 
which may be excluded on other grounds, but ignoring interrupts which makes it overly restrictive and hence a sub-LSCD), 
for the result that a run produces (if any). In addition there may be a sub-LSCD for the timing aspects of 
runs (which optimistically assumes perfect clocks and absolutely regular equipment).

\item The advantage of an SCFD over a mere LSCD that makes part of it  is that it provides a rationale for the technicalities of the 
central LSCD 
by explaining the interaction of instances with combinations of other system components. 
In this way the SCFD may provide a rationale for the 
technical ingredients of the central LSCD it contains.

\item An SCFD can be used as a rationale for its central LSCD. Clearly for rationalizing a single LSCD many different SCFD's 
can be imagined.  In principle their difference contributes to rather than diminishes the said rationale. 
One might require that an LSCD can be provided with a convincing number of extending SCFD's in order to 
demonstrate its plausibility. In this way the notion of an SCFD enters the process of LSCD engineering 
which is to some extent circular. This circularity is unproblematic because, as a piece of logic or mathematics, the 
LSCD can be given independently and in advance of further explanation (rationalization) in terms of one or more SCFD's.
\end{itemize}

\item{SCFD+IGUA:} {\em Incorporating an individual and group utility analysis.}
An SCFD can be augmented with individual and group utility analysis (IGUA). 
SCFD+IGUA is an extension (subclass) of SCFD in 
the following sense. For individual components (concept instances of the various concepts that the SCFD offers) 
as well as for groups of components various measures and degrees are defined which allow to express the utility 
that these components (groups of components) can assign to the behavior of other components or groups of components.

Only by means of an IGUA in addition to an SCFD
 it becomes possible to determine whether or not an individual component constitutes an 
optimal or near optimal solution to some formalized engineering problem. 

Examples:

\begin{itemize}
\item Software metrics can be used to asses properties of programs. Higher complexity (in metric terms) may lead to 
lower utility. Performance metrics can allow performance analysis. Both metrics and utility criteria may be included in an IGUA.
\item If a garbage collector features as a component in an SCFD (for a program execution architecture) 
 it may be included to express in an IGUA what 
advantage the running programs may have from the GC's presence. If  the system is multi-threaded there may be a 
vector of threads and the utility of the GC needs to be expressed with respect to the group as a whole. 
\item If within a distributed setting load balancing is applied the utility for each machine of the load balancing activity 
must be expressed. That may be done as a part of  an IGUA.
\item If a multiprocessing system runs a thread called ``virus scanner'' one expects an expression of the utility of this fact for 
other threads in the system (or at least for those threads that run so-called trusted code). For threads executing non-trusted 
code a formulation of negative utility is expected.
\item If a component is said to achieve authentication its security needs to be asserted in terms of non-interference with respect 
to the behavior of other system components. This involves complex definitions regarding collective behavior which we 
suppose to be placed in an IGUA.
\end{itemize}

\item{SCFP+IGUA+IGVA:} {\em Incorporating an individual and group value analysis.} A further extension augments SCFD plus IGUA with an individual and group value analysis (IGVA). 
This is an even more involved class of definitions. 
Now there may be agents around which are equipped with objectives and expectations as well as values and norms.

For instance in addition to a program there may be a programmer and the program captures the programmer's intuition. 
In addition to the machine there may be a user and the machine interface supports the user in achieving his or her goals.  
There may be another system user who assigns a positive value to an understanding of the system. Higher degrees of 
understanding may be valued more, this may include a positive valuation of the availability of proofs of system correctness. 
Other agents may consider a system in terms of trust, while not making an attempt to understand how it works. 
Trust is a value held by an agent with respect to some specific component. 
\end{description}

\subsection{Four classes within a continuous spectrum}
IDBR is rather open; it can be used to set the stage for a discussion and its specification suffices as soon as the participants in a 
discussion agree that that is the case. For the same concept many different IDBR's can be imagined. Typically books on 
programming will contain an IDBR for ``computer program'' and notoriously that IDBR will be presented as an 
explanation to the uninitiated which is better skipped by more experienced readers. 
Usually the very observation that disagreement is possible and that the explanation is a non-trivial 
exercise by all means is patently missing in works on computer programming.

The majority view in computing appears to be that those who have been programming by virtue of that particular 
experience know what it is (as an activity as well as in terms of what it delivers)
and that these so-called programmers do not need or appreciate any further reflection on what a program or programming 
might be. 
This is grotesque just as the judgement that those who
have been using money as a means of exchange or as a means of account or as a store of value or as any weighted combination 
of these functionalities would not profit from further reflection about definitions of money.

LSCD and SCFD are reasonably clear notions in the sense that whether or not an LSCD or an SCFD qualifies as such 
can be judged in objective terms. However, as soon as an LSCD is supposed to define a concept X for which an IDBR has 
been given already (or perhaps a plurality of IDBR's is around) deciding whether or not the given (proposed)
LSCD is appropriate for that very
concept is a wholly different matter. Obviously the question whether or not providing an LSCD makes any sense in a 
concrete case may also be debated. 

Examples from computing: 
\begin{itemize}
\item It is plausible  that `user' cannot be given an LSCD but it can be given a 
super-LSCD (often phrased in terms of an under-specification of user behavior) which may be useful as a part of an SCFD. 
\item Authors of books on computer programming never seem to appreciate the relevance of an LSCD for 
``computer program'' and often also don't bother to provide a sub-LSCD tailored to the specific program notation 
and programming environment on which their book has been based.
\item Middelburg \cite{Middelburg2010} claims that (in the published literature on computing systems) no 
LSCD for ``operating system'' can be found and that in addition no SCFD for
machine or for program contains either a sub-LSCD or a super-LSCD for ``operating system''.
\item For ``computer virus'' one finds high level IDBRs at best. Here we ignore very technical SCFDs based on 
rececursive function theory that  are so distant from programming practice that these may be classified as metaphorical
IDBRs rather than as SCFDs.
\item For the concept of a ``propositional statement'' LSCDs are far more common than IDBRs. 
\item The Turing machine is a common SCFD for ``computing device''. There seems to be a suggestion that for that reason it 
explains the concept of a program as well, which we cannot agree with.
\end{itemize}

Looking at the wider definition classes of SCFD+IGUA and SCFD+IGUA+IGVA it is obvious that 
demarcation is difficult and that we are 
looking at continuum of options of defining a concept rather than at stages within a discrete spectrum.

In the direction of more involved definitions one increases both technical 
comprehensiveness and philosophical completeness, whereas 
in the direction of LSCD one optimizes philosophical unambiguity and (extreme, perhaps even unrealistic) technical simplicity.

\subsection{Application of IDBR,  LSCD, and SCFD}

The following principles can be used to express the intended use of definitions.

Suppose one focuses on a family of related concepts for which an set of IDBR's has been given. These concepts are supposed to be 
simultaneously instantiated within some SUI (system under investigation). Further some specific mechanism or phenomenon taking 
place in 
SUI is taken in focus (The phenomenon to be explained or PTBE).

Now we define what an understanding of this phenomenon may amount to:
\begin{enumerate}
\item Each (or a significant subset) of the used concepts should be preferably provided with an LSCD.
\item For the whole system an SCFD should be given,

\item The PTBE should be defined either as part of the SCFD or in terms of an IGUA on top of the SCFD. 
Together these ingredients may be 
termed a model of the system under investigation. This should be done in such a way that 
\begin{itemize}
\item the occurrence of the 
phenomenon/mechnism depends only of properties of the components for which an LSCD has been provided, 
\item the occurrence of the phenomenon/mechanism should depend on properties of these 
components in a way which serves to understand the ``real system''. Now this is a circular requirement. 

\end{itemize}
By claiming a specific SCFD+IGUA to be relevant for explaining a phenomenon/mechanism these 
requirements are implicitly validated.

\item All pragmatic reasoning about PTBE and its occurrence inside the SUI which SUI users are expected to 
perform can be replaced by more 
formalized reasoning inside the model followed by an interpretation in the reality of SUI of the results of this model based reasoning.
\end{enumerate}

\subsubsection{Example: application to operating systems}
From \cite{Middelburg2010} we conclude that an LSCD (or an SCFD) definition of an operating 
system cannot be found in existing literature before 2010.
Providing an IDBR for operating systems is not a particularly exciting challenge as 
most books on computer systems provide a listing of services which an OS
is supposed to fulfill. We refer to \cite{Middelburg2010} for a listing of these classical objectives of an OS.

Interestingly there may be alternative IDBR's just as well. 
Here is an example: one imagines a class of machines which can be loaded with a range of 
programs. Now as it turns out all loaded machines have some functionality in common. 
Then this shared functionality may be extracted from each program
in the portfolio of programs.  Modified machines may already contain code for the shared functionality. 
Thus: an OS is a program that provides shared 
functionality for a class of program execution architectures. 
Exactly how this OS is combined with a loaded executable during execution is left untouched
just as the question how money may serve as a means of exchange is left unanswered by that particular IDBR of money.

Yet another IDBR for an OS is found by stating that an OS supports the run of a program by 
serving as an intermediary between the running executable and various
other system components such as IO and peripherals. This is a rather old fashioned IDBR. 
It is unsatisfactory because ``serving as an intermediary'' 
(or simply helping) is quite vague.

Given the preceding classification of definitions we are interested in finding LSCD's for ``operating system''. 
The structure of an LSCD for the concept of an OS may be as follows: 
\begin{enumerate}
\item An operating thread is defined as an interruptible thread (which can be provided with an 
LSCD in the style of \cite{BergstraMiddelburg2007}) 
that manages the execution of other threads in a polythreading environment.
\item An OS is the control code for an operating thread (see \cite{BergstraMiddelburg2009b} for an LSCD for control code, 
embedded in an SCFD that also provides 
machine functions).
\item The OS is supposed to be a program (for an LSCD we refer to \cite{BergstraLoots2002}) in addition 
to being classified as control code (see
\cite{BergstraMiddelburg2009b} for an SCFD containing a definition for control code).
\end{enumerate}

Given an LSCD 
for ``operating system'' one may extend it to an SCFD including LSCDs for ``program'', ``malicious code'', 
``executable code'', ``user'', ``operator'', ``configuration management''.

The phenomena that may be explained may include: bootstrapping, interrupt handling, garbage collection, multithreading and 
multiprocessing, program testing, program debugging.

\newpage
\section{IDBRs for money}\label{IDBR4M}
From the theory of money one may extract a range of different IDBRs for money. LSCDs 
can be provided for formalcoins, formalbanknotes, formal bank accounts and so on. More integrated pictures involve SCFDs that
take the interaction between various money classes into account. IGUAs will add an analysis of the usefulness  
of the different moneyclasses depicted in an SCFD. Including aspects of an IGUA adds information about large 
scale behavior of groups of users of instances of various money classes.

\subsection{Basic elements for an IDBR of money}
A possible IDBR for money is as follows: at any episode money (that is the family of then valid money classes) 
consists of any coherent category of entities, 
either abstract or  concrete, together with methods,  rules, or protocols of use which serve, in a dedicated fashion,
some or all of the following functions in decreasing order of importance 
\begin{enumerate}
\item \label{UoA} unit of account, 
\item store of value, (short term and long term to be distinguished; sign of wealth; money as an asset class),
\item means (medium) of exchange, (including: means of payment; means of settlement of debts),
\item legal tender (means of payment of taxes and fines; redemption of liabilities to the state),
\item standard of deferred payment,
\item standard of value,
\item \label{Dim} dimension (or factor of a composite  dimension) used for expressions that occur in formal texts about matters of organization,
\item optimum of liquidity (optimal readiness for serving as a means of exchange in a context of uncertainty about the future),
\item sign of political association, (usage of particular moneyage for one or more the roles listed above as a sign of loyalty),
\item means of communication (vending machine, booking systems),
\item barter killer (exclusive means of exchange in a state where barter is forbidden by law unless mediated by money 
for taxation purposes, law enforcement and surveillance; goods never buy goods see \cite{Davidson1972}),
\item a quantified right providing its holder with choice-value (\cite{Moini2001}), (alternatively: two sided balance sheet 
phenomenon see \cite{Bell2001})
\item a quantified right expressing its users' uncertainty (or worry) about future events,
\item commodity token, (including value-less commodity token; a physically independent, stable and easily transportable 
usually man made or selectively hand picked token representing a subset of the above roles in specified quantities).  
\item valuable commodity token, (a commodity token which possesses an independent 
and autonomous value for non-monetary purposes).
\end{enumerate}
The first three functions are mentioned throughout the literature on money. 
The fourth and fifth functions are often mentioned as well.

Rather than constituting a single IDBR this listing provides the components of a variety of 
IDBRs each requiring from the mechanism claimed to be money that it fulfills different subsets
of these functionalities and perhaps to different degrees.\footnote{%
This listing is by no means complete. In \cite{BergstraL2013}
an alternative listing of functions of monies is presented culminating in the concept of an exclusively informational money (EXIM).
A combinatorial explosion of different monies can be imagined; the listing of functions and qualities given above is rather
conventional and is not suited to fit new mines like Bitcoin.}

\subsubsection{A scheme of IDBR's for money}
The above listing provides a scheme for IDBR's of money, each IDBR
consisting of a weighted combination of the listed aspects (that is the basic elements), rather than a single one.

The number of possibilities is very large, even if all weights are taken either zero or one.
This possible proliferation explains why so many definitions of money can be found in existing literature.
We consider the combination of functionalities \ref{UoA} and \ref{Dim} as the core IDBR which all other 
IDBRs ought to include.

Further remarks:
\begin{itemize}
\item Implicit assumptions are made as follows: a means of exchange is to be understood 
from the point of view
of an unconstrained owner of the amount of money (used for exchange) or from the perspective
of its trading partner. For other agents related to either one of these parties the amount of money may not 
be such a means to themselves but they may only recognize that it serves as such for its owner. 

\item The different functionalities listed above are not orthogonal.  For instance the standard of deferred payment functionality may 
be considered subsumed in an appropriate combination of the preceding three functions. 
But in the case one or more of these three
roles are not fulfilled the standard of deferred payment functionality may yet be of independent importance, however.

\item Commodity money refers to all instances of money classes that meet the IDBR elements (value-less) commodity token or valuable commodity token. Commodity money class instances consist of material objects that might have a value of their own outside the
role played as pieces of money. Thus commodity money satisfies a basic IDBR combination that includes either valuable 
commodity token or (value-less) commodity token. 

Valued commodity money used to be very important but is now becoming outdated. Silver coins are a paradigmatic example of 
commodity money. Commodity money is a relational notion because it is in relation with individual 
other agents that the commodity that
a piece of money consists of has some value. This value is determined by way of exchange (at least in principle). Most if not all
commodity monies are used as money above par (i.e. forgetting about their moneyness results in a loss).

\item Perhaps only metaphorically it may be claimed that a unit of currency (say the Euro, EUR) is like a dimension 
in physics. Murat \cite{Murad1943}  concludes 
from that metaphor that money must be abstract, only a unit of account and no more. Assuming that viewpoint Euros don't exist 
as entities just as meters
don't have an independent (physical) existence. Euros are only a measure of value. 
For the concept of a dimension we refer to \cite{Page1952, Causey1969}.

\item Implicit in any specification of money is a community of agents which are its users in some form or another. 
Definitions of money that we have found
scarcely pay attention to the assumptions to be made about such communities. Often it is taken for granted that money using communities
roughly coincide with the inhabitants (and visitors) of one or more national states or substantial parts thereof
but there is no principled argument for that assumption.

\item We cannot provide an educated guess of the number of specifications of the concept of money
that can be found in the literature. It probably runs in the hundreds.  Although so many specifications of money can be found, 
the development of a systematic survey of these has not been either 
attempted or achieved. Most authors take their own favorite specification for granted or at best contrast
 their specification with one or two competing or preceding ones.
\end{itemize}

\subsubsection{Dimension and unit of account}
The roles as a dimension (\ref{Dim}) and as a unit of account (\ref{UoA}) are quite related. These together constitute a 
meaningful and
minimal IDBR for money.
This perspective is by no means new. We add the following remarks:
\begin{itemize}
\item
We consider money as unit of account (UoA-M) the most general IDBR of money. Ingham 2004 considers UoA-M a precondition
for MoE-M (means of exchange money) and traces that viewpoint back to Keynes at least; White \cite{White1984} states that a unit of 
account will remain 
wedded to a means of exchange. Keynes assumes that money both refers to an abstract UoA-M and to its 
implementation serving concurrently as as an MoE-M and as an SoV-M. David Laidler \cite{Laidler1997}, however, 
considers the means of exchange a sine qua non;
a viewpoint, which we consider less useful for our current objectives.

\item
Taking the UoA-M as a point of departure, any currency can be considered a dimension. A Euro is
comparable to a meter, a Eurocent is comparable to  a centimeter. Any money that can be used as a store of value can also be 
considered as a unit of account at a more abstract level, and the same holds for a money that is used as a means of exchange. Here we assume for 
simplicity that the different elements used for either storage or account are comparable of value in a linearly ordered fashion. 

\item Both means of exchange money (MoE-M) and money as a store of value (SoV-M) require some form of physical presence, which is certainly
not implied in the UoA-M concept of money. Neither function of money implies the simultaneous presence of the other function.
If $m$ is a MoE-M or a SoV-M then UoA($m$) denotes the corresponding unit of account money, which can be imagined in principle.
There is no implication that if $m$ is used as an MoE-M (or an SoV-M) that UoA($m$) is used as a money of account if any such money is used.
\end{itemize}

\subsection{Legal and relational IGVA elements for money}
In the preceding paragraphs we have discussed mechanisms for defining money that are derived from either the role of money or the 
mechanism it employs. The listing of IDBR elements for money is the basis from which a plurality of IDBR's can be composed.

But different aspects exist, such as listed below. In the context of our discussion of definitions of money these aspects may 
be considered IGVA elements. These
IGVA elements contain information or judgements about the legal and technical status of money classes 
with in their preferred financial system.
A complication that we will ignore is that when composing requirements for monies some subsets of requirements will produces
near--monies rather than monies.\footnote{%
In \cite{BergstraL2013} an attempt has been made to be very precise about the distinction between monies and near--monies 
when combining a heterogeneous collection of (near--money) requirements.}  

\begin{description}
\item{\bf{Fiat money.}} Fiat money consists of objects which are supposed to be representative of other forms of money 
that have a more primitive status. 

Banknotes issued by a commercial bank which guarantees at any time to exchange the banknotes for silver coins is a 
paradigmatic example of fiat money. The phrase fiat money expresses that trust provides essentially worthless items with
value in the same way as commodity money already has.

With commodity money becoming outdated (as a technology) the same holds for fiat money (as a notion) 
because of the contrast losing its relevance.

\item {\bf{Fiduciary money.}} The same as fiat money, sometimes only bank money (which includes fiat money).

\item{\bf{Chartal money.}} Chartal money is money which derives its value from an agreement (a relational notion).

According to some authors chartal money can exist without a state though some hold that (chartal) money can only exist if it is 
backed by a state. 

\item{\bf{Legal tender.}} Legal tender money is money which the state will accept at its tax offices. 

Visiting a tax office for the purpose
of tax payment is becoming rare and for that reason the tax office is supposed to include the bank accounts used by the tax office.
Commodity monies and fiat monies are usually legal tender. In practice, however,  only bank account money is legal tender, however,
as tax offices prefer bank transfers for tax payments.

\item{\bf{State money.}} State money is chartal legal tender.

\item{\bf{Base money.}} Base money is a phrase that must be understood in the context of a financial system with private households
(including firms), commercial banks and a central bank. Then base money combines the currency (commodity money plus fiat money)
owned by households and banks with the deposits held by banks with the central bank

\item {\bf{Financial derivatives.}} Financial derivatives bridge the gap between the capital markets (and commodity markets) and 
money as a technical device helpful for organizing the economy. Derivatives define money in terms of uncertainty and 
competition rather than merely to provide tools for dealing with these phenomena. 

\item{\bf{Informational money.}} Informational money has money classes with informational items as 
instances only.\footnote{%
In  \cite{BergstraMiddelburg2008}  a financial account is specified in terms  of 
process algebra, and in addition the point is made that money in its 
future manifestations might become a computational phenomenon altogether, 
and more specifically a topic crucially depending on 
computer system specification and verification. Indeed it is conceivable that from some stage onwards 
money consists merely of data, or perhaps data encapsulated within
appropriate protocols. This would be in accordance with the Jevons' interpretation of Gresham's law.}

\item{\bf{Exclusively informational money.}} In 
\cite{BergstraL2013} exclusively informational money (EXIM) is introduced as informational money for which 
access takes priority over ownership. A non-exclusively informational money (called technically informational money and abbreviated TIM) provides a holder of informational items serving as instances of its money classes with some rights that exceed the 
mere consequences of having certain information (such as a secret key) at his disposal.

\item {\bf{Virtual money.}} With virtual money we will refer to an informational money that has been defined on top of and interns of another (possibly also informational) money. The bookkeeping system of a virtual money
suggests the existence of accounts (like bank accounts) and offers operations
to manipulate these accounts but it performs no more than
a visualization (conceptual organization) of sums of money owned by an organization that in reality are organized differently. This 
use of the term virtual has been borrowed from computer science where virtual memory, virtual machine, and virtual network are
commonplace.\footnote{%
We don't mean to say that electronic 
money is virtual. We also don't mean that virtual money is a money of a virtual world, although it could be.
Bank accounts with electronic access are non-virtual. Virtuality results if on top of an ordinary system of bank 
accounts a quite different (usually much more detailed) organization of accounts and transfers  is implemented which is 
customized to particular objectives.}
\end{description}

Below we will discuss agent role based explanations of money. For SAR (subordinate administrative role) we will provide a detailed 
description of a role based ``theory of money''. 
In terms of the above classification P in role SAR needs to acknowledge only fiduciary money
in electronic form. The view of money of  SAR as depicted below constructs an ad hoc virtual money tailored to his needs and circumstances.

\subsection{Further dimensions: institutional perspective and metaphysical status}
It may be maintained that money is at best an element of a financial system which needs to contain other elements in
order to serve its purposes. When posing the question ``what is money'' an attempt is made to abstract from the institutional
dimension and to single out that one element while forgetting other aspects. Whether or not such an abstraction can be
made in a meaningful way is hard to judge in the absence of a theory of abstraction which may apply at this level of generality.
Some description of what an institutional perspective on money might offer is  given below.

Another aspect which must be contemplated when defining money is its metaphysical status: is it real or is it an 
epistemological notion which for its existence and meaning is fully dependent on what people write, say, and think. Below
a commitment is asserted to a realistic perspective on the metaphysical status of money.

\subsubsection{Functional perspective versus institutional perspective}\label{MoneyagePreference}
The scheme of definitions outlined above is well-suited to explain different monies and financial institutions. However, as 
Merton \cite{Merton1995} points out it may be preferable to think in terms of financial functionalities rather than in terms of financial 
institutions. Merton lists six functionalities which a financial system needs to provide. His listing may be considered an improved
version of the usual listing of roles of money, with the difference that an appropriate financial system needs to fulfill each of the six functionalities.

Attempts to settle the question what is money might profit from making an underlying assumption that an adequate financial 
system as explained by Merton exists. In that case the special role of money  may become less unique. One might ask for 
instance: what is 
the minimal role that money can play in an adequate financial system. Let moneyage preference be a measure of the 
degree to which
a society is willing to accept the burden of having all or most transactions coupled with payments from a moneyage in 
a financial system. Now
new questions can be formulated: are fluctuations in moneyage preference observable and if so, can such fluctuations 
be used for policy 
making purposes. This is not about the desirability of free markets, a market can be free on the basis of barter. 
It is rather about the
degree of barter supported by and required for the proper mechanics of a particular financial system.

When taking this path IDBR definitions gain importance at the cost of bottom up LSCD (SCFD etc.) definitions.

\subsubsection{Money: is it real?}
Many authors suggest that money exists at some distance from the real economy. Keynes is seen as someone who had an
opposite view. Novel financial instruments have been blamed for missing contact with the real or underlying economy and
regulation is seen as an instrument that can prevent money from becoming an autonomous phenomenon. 

Few will deny the realistic status of a golden coin. But if money is a `two sided balance operation' resulting from the simultaneous
creation of credit and debt, expressed in transferable IOY's and waiting for its unavoidable annihilation, then an epistemological 
approach to its understanding may be preferable. So where is money on this scale, or where are various money
classes on this scale. 

A realistic view seems to advantageous and defensible for well-known money classes with wide acceptance. But monies in
deviating financial systems, for instance the virtual monies that will be the focus of our final section, may best be understood 
with a constructivist perspective resulting in a primarily epistemological status.

Insisting on a realist perspective concerning the metaphysical status of money has consequences for the interpretation of definitions.
Under a realist interpretation definitions are specifications against which observed phenomena are matched. 
The focus on imaginative definitions,
as emphasized in this section seems to constitute a commitment to epistemology or to a constructivist approach towards 
the metaphysics of money. It is
not meant that way. Imaginative definitions are supposed to be helpful by producing mental constructions by means of
which a confrontation with real phenomena is to be facilitated. It is not meant in any way that these mental constructions (images)
can or should  replace parts of reality. But in as far as reality is created by ongoing human design such constructions
may constitute, in principle at least, a useful tool for that design activity.

Some clarification for these contrasting perspectives on the metaphysical status of money results from comparing this theme with 
the topic of risk. A fundamental question on risk is whether the metaphysical status of risk is to be understood from a realist 
perspective or rather from an epistemological perspective. Rosa \cite{Rosa1998} insists that risk is a realistic phenomenon. 
A risk exists objectively and may or may not become aware to those at risk, who only if that happens, or at least only
after they have developed a suspicion of that risk,\footnote{The suspicion of the risk
not being any confirmation of its existence, however.} can 
analyze the risk and determine and effectuate a policy towards it. Thus, by merely talking about the suspicion that the risk exists that 
the sea level rises 250 meter in the coming 10 years no such risk can be constructed, irrespectively of how many persons 
participate in the debate. Some scenario that leads to that rise
of the sea level should be provided to make the case of the existence of the risk. Of course the risk can exist if no such scenario is 
known, but its existence is not made more plausible, let alone created, by its mere contemplation. One may speak of a 
potential risk, this being a concept from epistemology. Having acknowledged the potential risk as a risk (which requires the
demonstration of a scenario)  still embedded in
uncertainty without any further quantification or qualification, it can subsequently be analyzed and
perhaps probabilities can be assigned to it which can form the basis of policy and action. If no probabilities can be assigned,
even by sustained scientific activity, and if 
in addition human action could in principle avoid significant adverse consequences if the risk materializes, 
such action may be prescribed (or somehow justified) by a policy based on the precautionary principle.

The comparison with risk is meaningful for reflection upon the metaphysical status of money because it demonstrates that
a realist position can be successfully maintained in combination with epistemological perspectives on related concepts. 
This seems to be a useful perspective concerning the status of various money classes and financial systems as well.

\newpage
\section{Agent role dependent views of money}
It is a remarkable feature of the vast amount
of ToM\&F that so many viewpoints have been developed and mutually confronted without an equally strong insistence to express
or at least analyze in which circumstances each of these paradigms is most at its place. 

So we assume that moneylike phenomena are present in varying degrees in various circumstances. Indeed astronauts on the 
moon around 1970 
have not  paid there with money in the same way as museum visitors  used to in those days. It can be safely concluded that
money in its many manifestations can be present in a variety of degrees, depending on place and time. 
So one can imagine different scenarios each of which are best
specified and explained with a particular subset of the paradigms that have emerged in theories of money. This leads us to
contemplating agent role dependent perspectives on money.

\subsection{Five agent roles and corresponding observer roles}
More interestingly, even at a fixed time, his most profitable perspective on money may well vary 
depending on an agent's role in society. 
The theory of money that describes financial matters in a most useful fashion for some agent or group of agents 
need not be unique for its historic episode and economic system. On the contrary one may assume some role or 
collection of roles and
make an attempt to design a perspective on money most appropriate for those roles. Rather than to acquire an abstract understanding 
from a broad, scholarly and impartial perspective, research may be aimed at formulating a pragmatic theory which has an optimal
explanatory value for a moderate range of phenomena each of which matter to the researcher  the roles which determine the 
perspective that a researcher has chosen as his point of departure. 

Here we describe four roles while choosing the first one (SAR) as the preferred angle.

\subsubsection{The subordinate administrative role (SAR)}
SAR includes management of operational organizations of up to say 250 employees, as long as no large investment 
decisions need to be taken.
Moving to a new building, upgrading the network, downsizing PR, hiring new staff, executing staff reductions are all 
part and parcel of the SAR role. Most instances of
SAR are found as lower middle management in larger companies or institutions.

The SAR is characterized by the following aspects, formulated from the perspective of a person P in the role SAR:
\begin{itemize}
\item If P has difficulties with comprehending phenomena of money, that relates to one of the following items:
\begin{enumerate}
\item taxation: this poses optimization problems for which P is ill-equipped.
\item investment and financial management: P avails of no explicit strategy or methodology,
\item insurance: not even a minimal insight in the cost calculation of insurances is available to P.
\item Technical complications with (internet) banking: what security levels are needed, how to guarantee those; 
what information may
be required from one's financial agent at this bank, and what information will this person leave undisclosed.
\item bookkeeping complications: on the workplace P may be confronted with very complex methods for and automated systems of
budgeting and financial planning. A lack of understanding concerning day to day financial practice in P's professional environment may 
exist either consciously or unconsciously. 
P is told that this lack of understanding is to be remedied by means of courses and training, but that is
plain nonsense (though P cannot afford to say so).
\item Inability to handle the automated systems for distributed account management as prescribed at the workplace.
\item When endowed with some financial responsibility: inability to carry this weight in a solid fashion, 
both in terms of financial techniques and conceptually, often aggravated by a defective grasp of the financial context and the fit
in an overall picture.
\end{enumerate}

\item The mere conception that an abstract or general
 theory of money would matter is considered flawed by P's colleagues and superiors. It is taken for granted that no
systematic insight in the nature of money is needed to master the difficulties encountered.

\item P perceives a lack of money resulting from either defective income or from overspending, or from 
disappointing investment or any other form of bad luck.

\item The need to support other people (children, partner, parents, or friends) who in fact have even less clue about money to the extent of 
being afraid of dealing with it in whatsoever role except spending money for (almost) immediate consumption.
\item Lack of orientation experienced by P when the local, regional or national political system system asks for a democratic vote: 
how can P balance the different stories put forward by candidates from different signatures about national budget cuts, state 
overspending consumer debt menace, exploding cost of healthcare, and so on.
\end{itemize}
Thinking in terms of our classification of definitions, this listing completes an IDBR of SAR for which an LSCD cannot be 
provided in a plausible manner.

\subsubsection{The SAR observer role}
For an observer Q of agent P in role SAR it matters which aspect of P's behavior is investigated. The description of 
SAR has been designed
so as to minimize his required understanding of money outside his job. Q may take an interest in how P deals with money
outside his job, because that may be influenced in non-trivial ways by his dealing with money within the job. More 
plausible, however, is that Q takes an interest in how P deals with money inside his job.

Below we will provide a detailed account of a theory of money for P. Q, however, must take into account the 
possibility that P's
theory of money is problematic, unhelpful, or simply incorrect. For Q to be able to investigate whether or not P in fact
believes some specific theory of money (and to assess to what extent such a belief is an asset for P) 
requires a full awareness of money outside and inside the organization where P is employed.

\subsubsection{The information analyst role (IAR)}
IAR, the information analyst role,  has other professionally related issues to deal with than SAR, whereas the private life related perceptions of money are 
similar to those of the person in an SAR role.
\begin{itemize}
\item The need to grasp all forms of financial data and processes that need to be distinguished in a medium size enterprise.
\item The urge to understand the validity of complex management accounting systems.
\item The need to understand the principles and procedures of financial auditing, in particular in the context of highly automated bookkeeping.
\item The need to communicate about these matters with persons who combine a deplorable 
lack of knowledge and interest in computer systems with a similar disinterest in finance, 
often failing to see that these matters are far more complex than
the average subject that these individuals have to deal with in their own daily routines.
\end{itemize}

\subsubsection{The IAR observer role}
An observer of IAR agents like an observer of SAR agents must be able to combine a theory of money with a theory of
accounting. He also must be able to imagine dedicated theories of money and accounting which an IAR agent makes 
up for himself. In addition this observer needs to understand financial information systems in general and the system
used by the agents he is observing in particular. This is a challenging task for which no ready made theoretical 
preparation seems to be available.

\subsubsection{The general management role (GMR)}
\begin{itemize}
\item Besides being responsible for professionals of varying kind, the GMR role brings with it the responsibility to directly manage 
employees in the SAR and the IAR roles.
\item GMR often needs an understanding of business economics: how to finance a new building, how to use financial engineering for risk
management, how to report to external authorities.
\item In the case that P when acting in role GMR is asked to make significant investment decisions a 
perspective on the future regional economic development may be required.
\end{itemize}

\subsubsection{The GMR observer role}
The GMR observer role imposes the same qualifications of financial theory awareness
on an agent as the GMR role itself.

\subsubsection{The housekeeping role (HKR)}
For P in HKR we expect that:
\begin{itemize}
\item P needs to have the ability to develop the households balance as 
well as its profit and loss statement in as far as that is needed to 
fill in the tax forms. This requires very modest bookkeeping skills that can usually be done by hand.
\item P will use money as an MoE when shopping.
\item P will never write documents about money.
\item P will know that money can serve as a store of value. 
P will not perceive money as an asset class surrounded by alternative means
of investment.
\end{itemize}

\subsubsection{The HKR observer role}
The housekeeping role (HKR) can be characterized as a simple version of SAR, which 
is in fact so much simpler that the
move towards a formalistic position which will be advocated for P in role SAR,
fails to have a convincing or even decisive incentive for P in role HKR. P in role HKR will 
not distinguish between Euros
and Formaleuros. Housekeeping P will probably only deal with a limited number of money classes from the 
current money class family (including near-monies).

Although for P the reward of a formalist approach may be absent, a spectator of P say Q, for instance 
an ergonomic analyst or consultant of 
housekeeping
activities may have a different perspective. Here are some questions that Q might pose.

\begin{enumerate}
\item How is P taking care of money-items, both coins and paper, overnight and in weekends. Is P ever 
transporting money within his house. Is there 
a distinction in the security of different locations for storage. 
\item How many persons of P's household have access to the currency that is stored at any time.
\item How long will P keep currency items in store, on average and maximum length.
\item Is P aware of all locations of currency in his house at all times. Has P recently been searching 
for lost currency in house. Has P recently 
discovered money that he had forgotten. Has it occurred in the last year that P has lost and 
rediscovered the same coins or banknotes twice of even more.
\item Is P rational in his decisions to draw currency from a bank account. 
\item Is P ever returning cash to his bank account. Is he ever returning cash in the form of coins. 
Is P saving money in order to return it to his bank account in a 
later stage. If so is P exchanging the saved money for coins and banknotes of higher 
denominations or are all coins and banknotes saved until 
being taken to the bank.
\item Does P have a preference for the use of specific coins or banknotes.
\item Is P maintaining a bookkeeping of its holdings of currency, if so how often is it recomputed.
\item Is there any rationale for the amount of cash that P has in stock at home on average.
\end{enumerate}

Several of these questions require that different coins of the same coin type (kind) version are distinguished. If these 
questions are to be taken seriously formal 
models have to be developed about the location of amounts of money in P's home at various moments of time. 
Sometimes it is necessary to refer to
the counterpart in a model of a physical coin or banknote. Then it is quite useful to work in a model (and for 
that reason to make use of Formalcoins)
rather than to discuss real coins and banknotes. The problem with the latter is the open ended character 
that results from the assumption
that real and material objects are dealt with. All questions mentioned above need to be cast and if possible 
answered in a formalized model.

\subsubsection{The point of sales executive (PSER)}
One step further down the scale (if this implicit judgement of value is at all legitimate) one  imagines a 
person P who sells ice-cream for 
currency and who is employed by a firm for a fixed amount per day while being granted a fixed 
percentage (say 80\%) of  the tips he receives.
These tips will be paid as a part of the income by means of a bank transfer.
This role is a point of sales executive role (PSER).
P is specialized in currency based transactions where he is always the seller. 
Thus for P acting in role PSER, money is indeed a means of exchange though merely in one direction.  
The money of account function is rather minimal. 
It is unlikely that P (in role PSER) needs more than a naive form of financial 
realism to manage his own actions. 

\subsubsection{The PSER observer role}
Now we envisage an observer Q of P and his colleagues. Q is specialized in observing PSER personnel. 
Q acts as an applied psychologist
and consults the employer of P about who to hire next year.
Q may be served well with a formalistic approach where he makes use of the notion of a Formalcoin for 
each of the
values offered by the Euro coinage. Indeed Q may pose a range of questions which increasingly call for a 
clear temporal and spatial picture of the activities of P and other PSER operators.
Q may want to use formalbanknotes when expressing that P has announced that he will not accept
banknotes above 50 Euro for certain purposes. Here Q speaks of P speaking of his handling of types of currency
items, and this level of indirection may be profitably represented in the type system that is used.

Many questions about the mechanics of token handling can be posed.  
More often than not the issues raised are independent of the moneyness of
the tokens involved. That provides a reason for preferring relevant process descriptions in 
terms of formalcoins rather than in terms of coins from the Euro coinage. 

Here we provide an extensive though unsystematic listing of such issues. It should convince the reader that the PSER 
observer may
profit from thinking in terms of formaltokens (= formacoins and formalbanknotes). These questions are not 
considered research questions on money
(or the use of coinage) per se. Instead these questions indicate the level of detail which is needed for the
analysis on systems within computer science. A formal theory of money is needed in order to make progress on such
questions at all.\footnote{Realistically speaking this is rather unlikely to
happen given our assumption that coins as commodity money have had their longest time 
and will soon be replaced by electronic media
thereby rendering questions as mentioned below rather futile.}

\begin{enumerate}
\item How much currency has P available during his working hours. 
How is this distributed over various containers of currency items. Is there an explicit strategy
for storage of coins and banknotes, in particular by means of sorting the tokens according to their value. 
Is this strategy taken from employers directives or has it instead  been defined by P on his own initiative. 
Is the fraction of banknotes
increasing during the day. Are banknotes with high denomination stored in a different way. 
How is the separation between tips and payments made.

\item Is the bookkeeping of tips physical, 
that is a separate location (wallet or wallets) is used for tips, or is some form of bookkeeping used. If a 
separate physical storage of tips is used, is there any connection between coins and banknotes that end up in that 
store and the way in which the respective coins and banknotes have been offered to P by his customers.

\item What strategy is P using to maximize the chance that during the day he will be able to provide return money
 if customers want to pay with
amounts that exceed the price of what they buy with less than the value of the lowest token that they offer. 
If P needs to provide change, how will he
retrieve it from his various money stores. 
Does P have different strategies for making an assessment about whether he can find change and 
actually looking for the change.

\item How often is a purchase failed because P cannot offer change. 
How many false negatives are produced (wrong judgements of inability to
produce change). Is it the case that returning false negatives are a preferable strategy in some cases 
(indeed if P has to hand over near all his small
tokens in order to accommodate a minor purchase which a customer intends to pay for by 
means of an expensive paper  token this action may
prove quite counterproductive for P). Is the strategy P has for determining a (physical)
 amount by which to return change stable or is it somehow randomized. 
 
 \item Are there cases where P has and expresses a preference for a specific way in which 
 a customer composes a payment as an amount with precisely the value due.
 
 \item Are their cases where P prefers an inexact payment (that is one that requires change) 
 over a payment of the exact amount due. If that is the case, did
 P make up his mind about that state of affairs well in advance or did he make that judgement on the fly. 
 
 \item Is P ever making an appeal to other customers in order to help out with failed attempts to provide change, 
 for instance by asking the customer first to
 perform a change with another customer by himself and then to pay with more appropriate tokens.
 
 \item Is P ever asking for help by third parties in the case that a failure to provide change constitutes a false negative. 
 And if so, is that a conscious decision, or is it merely an oversight.
 
 \item Instead of refusing  to sell a product P because of unwillingness to provide change may also 
 express a preference for another amount to
 be offered by the customer. How often does this take place and can one assess that doing so indeed proves an advantage for P.
 
 \item Is P offering change in such a way that customer preferences are taken into account. 
 For instance if a particular money class is useful
 for paid parking. Or for buying a ticket for public transportation, or (almost outdated) for making a phone call.
 
 \item How often is P involved in a conflict where a customer claims not to have been given due change.
 
 \item Is P at all concerned about the validity of the tokens he receives. If so what definition of a correct (or valid) 
 token which he employs. What checks are performed at the time of receiving payments. 
 What action is taken if P finds out that a token that he 
 receives is invalid. Will he hold that against the person who offered the
 token and how will he do so. Is it ever the case that validity of a token is successfully challenged after a deal has been completed. If 
so what actions are undertaken.

\item Are there any cases where P can make use of information about which recent customer has handed over a specific token. 
Is such information maintained in any systematic way if at all. 

\item To what extent is P dealing with different customers concurrently. Is P used to obtain payment before deliver, 
or the other way around. 
What determines P's preferences for this matter and how does P enforce (if at all) his preferences on his customers.

\item If P asks for payment in advance. Is a bookkeeping made of that. Is P ever forgetful of the fact that he has already 
received money.
If he needs to provide change, is tis done before or after delivery of the product. If both orders occur what determines 
which choice of order is actually made.

\item To What extent is P's activity deterministic. Is it less deterministic if the identity of tokens is taken into account. 
 
\item How often on a single day is P handling the same coin. Is P ever handling the same paper token twice on a single day. 

\item Is the token handling behavior of P dependent on other circumstances than the sequential ordering of customer requests
that he receives. In particular is P performing rearrangements and optimizations of his token stock 
when there is a temporary absence of 
customers. Is P returning change in a different way if he observes a queue of customers.

\item How to measure P's performance with respect to the handling of tokens. Are optimal strategies available. 
Are such strategies dependent on
any classification of circumstances or market conditions.
\end{enumerate}

\subsection{SAR in more detail}
Because of its abundant proliferation we will now  focus our attention to the SAR role. The objective is to determine 
an appropriate level of abstraction for our considerations.
We start with listing some requirements on an SAR's dedicated theory of money.
\begin{enumerate}
\item SAR needs an abstraction of the theory of money. That means that certain aspects and features 
can and probably ought to be
hidden while other aspects are highlighted. In the computer science tradition of this use of abstraction 
there is no implication that only minor details 
are made invisible or even that what remains is a faithful representation of the original system or structure. 
The required abstraction in this case 
provides a conceptual interface that allows generic person P in role SAR to deal with money in an appropriate way.
\item Rather than following Hicks (money is what money does)  P believes that money is what money does for P.
\item After appropriate abstraction P has available to him a theory of money, which is probably (and even 
intentionally) incomplete by being unable to deal with features and phenomena outside P's scope. 

\item It is reasonable to make an attempt classify this abstract theory in terms of the dilemma's and 
contrasts that have been 
developed in ToM\&F. If P's reduced (and for that reason abstract) view of money for example happens 
to be fully explained by means of commodity 
money where the appropriate commodity is gold that 
can be delivered and stored in arbitrary 
weights, that is fine. The mere observation that this particular form of metallism is outdated and for that reason 
fails to explain the overall financial system before P's particular 
abstraction was made in this case, is immaterial. P needs (is entitled to, looks for) the simplest theory of 
money that serves his purposes.

\item The theory of money need not (but might profitably) be sufficiently 
expressive to tell the story of the history
of the financial system and related procedures which P is supposed to make use of. However, it may be very useful if 
the abstraction that is made use of
covers so much ground that comparison with similar organizations or units in the same and other 
(but similar) enterprises can be made or at least
understood by our target employee P.

\item Specific for SAR it is fair to assume that P need not be aware of the following matters in finance at least: 
\begin{enumerate}
\item how banks interact with central banks, 
\item how and why quantitative easing and other practices that increase central bank assets are applied,
\item how interest rates come about, 
\item how to count the total amount of money, 
\item how to define the circulation velocity of money, 
\item how often coins and banknotes are renewed and for what reason, 
\item how to detect counterfeit coins and banknotes, 
\item how, why and when to defend the rate of the Euro against other currencies, 
\item how to comply with international regulations on banking,  
\item how to measure GDP and national debt in order to monitor global key figures, 
\item how to avoid failures of individual banks, 
\item how to design the financial architecture of an internationally operating company, 
\item how to determine risk assessments and appropriate pricing for stock, bonds, options, warrants, 
swaps, and more sophisticated products of financial engineering, and so on. 
\end{enumerate}
\item All these high level matters surface at once if P is casting his vote in national elections. 
Now he is entitled to follow the advice of his favorite politicians
not requiring himself to understand the technical background of their positions. 
This also applies if the top management of P's employer
explains the incumbent securitization of buildings which were simply owned until recently, or if the necessity for and technical details
of insurance policies against higher interest rates are expounded which are supposedly needed when new premises are to be built.
\end{enumerate}

\subsubsection{Why money might be an issue for SAR}
One might state that at the level of SAR money is trivial. 
That this is not so follows trivially from the observation that the financial world which P 
has to deal with is not significantly less complex (and in many respect much more complex) 
than the financial world of say The Netherlands as a whole some 200 years ago.

One may object that most colleagues of P don't pay any attention to this kind of reflection. 
They seem not to be in need of an appropriate level
of abstraction and a corresponding theory of money and finance that serves them consistently 
through an extensive series of activities and decisions.
This disinterest in a theory of money that makes one's own life simpler and better explained is a topic for anthropological or perhaps 
sociological research. 
Our restriction to SAR is not meant to create a pragmatic application perspective which the
serves as a justification for the work to be validated by demonstrating that 
SAR employes informed by the outcome of this research perform better.
It may well be that psychological factors outweigh the effects of this kind of conceivable enforcement. 

\subsubsection{A  provisional and formalist theory of money for SAR}
We will assume that in his household P has a partner willing to perform basic financial processing where P acts merely as a 
consultant. P's handling of money and near--monies is very limited and  requirements on a theory of money 
that come about from money handling are minimal for that reason. 
Having thus simplified P's usage of money here is an informal and provisional theory of money for P:

\begin{enumerate}
\item All money is informational, digital, and electronic, 
it exists on chip cards and credit (debit) cards and it resides in ATM's and so on. If coins or 
banknotes are needed these 
are received or extracted from ATM's, or collected from previous uses just to be used for further transactions. 
The arithmetic of coins and 
banknotes is hardly important as for most transactions the form of payment is rather standard. 
You don't pay coffee with 100 E and so on. 
Cash management
in a restaurant is more concerned with adequate delivery of a tip than with 
calculation let alone with finding optimal  combinations of coins 
and banknotes. Here P may seek to optimize the likelihood that change can be returned, 
or that a next payment can also be made, or the weight of
his wallet after transaction, or even of deliberately getting rid of forged coins or of false banknotes. 
Cash management in shops is often made redundant by using specialized information technology, nowadays consisting
of apps that work on a general purpose smartphone or tablet. 

\item In normal circumstances no count of money is ever performed by P or on behalf of P. 
Shortage of money rarely takes place and it is signaled by refusals at pay stations where it merely 
implies that some preparatory transfer actions need yet to be accomplished. If that fails something `is wrong with the system'. The 
explanation is probably not about money at all. Coins and banknotes are merely
perceived as tickets providing admission to various activities. 
These tickets can be bought and payment is performed electronically. That coins and
banknotes are intrinsically money whereas a credit card is not is considered confusing. 
Coins and banknotes simply have costs like tickets
for a museum or for public transportation. 
No admission is made that currency is a most plausible (if not the most plausible) form of money. 
Important items are never bought by means of an exchange of currency.

\item Technically speaking P lives as if for daily life there will always be enough electronic cash available. 
This he concludes from the sociology
of his life style in combination with the architecture of his personal financial system as agreed with his partner. 
In other words, P deals with cash as with fuel. If it is out it needs a refill. 
The very thought that fuel might have an interest rate or that it might be used for very different purposes (like
buying a house) is considered absurd. 

\item As far as P knows electronic money is destroyed upon consumption (usage). 
P will never assume that money spent in payment 
readily comes available to a trade partner for whatever purpose. 
This is an obvious matter: what is spent as a daily expense is collected as income, a quite different type.

\item In as far as P has trust concerning the use of his money, P places trust in entire technology chains. 
Lack of money is a marginal problem
which is overshadowed by all other ways in which its implementation can fail to deliver its intended service.

\item P acknowledges a single generic and dominant interface for dealing with money: 
exchanging texts (including multipage spreadsheets) that speak about money 
(at least contain many figures that should be understood as carrying a unit of money (even if in practice the usually don't)
in relation to matters that are important for P, sometimes for his household though mostly for his employer. 

Money occurs in a text always in tables or formula's as a rational scalar followed by the unit
Euro. Now P has not the slightest conception of Euros, 
and always when a text speaking of money arrives in his mailbox he makes a (mental translation) and reads FE (Formaleuro) for 
Euro.  By means of this single action, which constitutes an abstraction by itself, 
all knowledge that he might have about Euros either directly or obtained from family, friends or colleagues
ceases to be applicable when evaluating this text. 
P will reply with similar texts using FE until the all but last stage and then cast FE as Euro, while 
hoping that the additional meaning thus acquired is neither confusing nor misleading for his colleagues, family, or friends.

\item Only by abstracting Euro to FE, P can ensure that he has full control about the assumptions that 
lead to conclusions or even decisions. 
Assumptions about FE cannot change (by definition) without P being aware of it. 

The philosophical status of P's perspective on finance may be something that P cannot self diagnose. 
This requires a philosophical 
awareness and sophistication he may not have. The following observations (both about P and by P) can be made nevertheless.
\begin{enumerate} 
\item For P it is an open matter whether he needs to share his FE
based knowledge with other individuals  in a similar position. 
Probably by doing so his grip on the subject increases, but very few people he
knows accept the shocking abstraction implicit in moving from Euro to Formaleuro (FE) as the dimension of the money of account.

\item By working in terms of FE's rather than Euros P subscribes a formalist position towards money, 
to be contrasted with financial realism, comparable to formalist positions in the 
foundation of mathematics. P is a subjective financial formalist. 
He operates in formalist mode irrespective of others who may well operate in a
realist mode an who may have sound justifications for so doing. 

\item An even higher degree of intellectual maturity is obtained in principle, 
but for P unachievable and for that reason devoid of pragmatic importance, 
if P operates as a financial intuitionist. That requires having available a constructive theory
of all designs that underly the various texts which P confronts, and not ever to rely on 
classical inferences from assumptions which themselves 
have not been understood in a constructive manner. 
(Whether an intuitionistic approach to finance can and/or should 
be distinguished from a constructivist approach is unclear to P and remains to be seen.)

\item For a selected subfamily of the current money class family (CMTF see item \ref{Cmcf} 
in subsection \ref{SubsectPROM} of Section \ref{SecTIMF}) P will introduce FE counterparts. 
Probably money classes that are specific to large scale international finance, and mainly used by banks, 
hedge funds, investment funds, and national and international financial authorities will be omitted, while
 the bookkeeping system that P makes use of may provide a plausible incentive for the
introduction of additional money classes outside CMTF in order to model various kinds of 
accounts that the accounting software has on offer.

\end{enumerate}

\item P follows Murad \cite{Murad1943} in the insistence that money is merely a dimension, and not a physical entity. 
Documents containing expressions for quantities with dimension 
FE are most common in the texts (documents) that P needs to read or write.
But other units occur as well. For instance: TIME (measured in seconds or hours, days, months or years), 
PRODUCT (measured in units of
a specific product, e.g. Chair of type $t_1$, Table of type $t_2$, Cupboard of type $t_3$ and so on), 
SERVICE (provided or consumed) can be measured in 
numbers with the unit specific for the service. For instance SERVICE: International mobile phone connection (say abbreviated
in an ad hoc style to IMPC). The price of IMPC may be expressed with a dimension of FE/SEC. 
This indicates that composed dimensions with FE as a factor are plausible. 
The performance PERF($p$) of an IMPC provider $p$ may be expressed with a quantity of dimension SEC/FE, with 
PERF($p$) = $q$ SEC/FE  expressing that $p$ provides to each of its customers 
 $q$ seconds of IMPC in exchange of one FE. Managers of the IMPC provider $p$ may feel the need
to carry out an investment program in order to improve the performance. 
A calculation may lead to the conclusion that with an investment of $k$ FE
all of its user base can be provided an additional $q_a$ seconds of IMPC. 
Assuming that the management of $p$ asks technical installations provider
$r$ for the implementation of this improvement (to which it agrees against cost $c$) then 
$r$ can be said to provide an improvement process with an 
effectiveness that can be quantified as $\frac{1}{c}\cdot q_a$ SEC/FE$^2$. 

For P the occurrence of powers of FE in composite dimensions is as plausible in 
principle as the occurrence of powers of SEC in composite dimensions in mechanics. 
Which composed dimensions are actually meaningful is another matter of course. 
P need not infer from these considerations that Euro$^2$ is a dimension
to be expected in practice as well. Nevertheless P may have a use for powers of FE in transformed documents (originally 
using composed dimensions with only a single power of Euro) that he expects to employ for further formal analysis.

\item Rather than that amounts of money serve as a means of exchange texts about money serve as a means of exchange. This equips P with 
a document based perspective on financial exchange. This can be further illustrated by means of an example. P  
considers buying a home and taking a mortgage a document based purchase. We assume that
 he buys from scratch paying by means of money transfer 25\% as an 
amount which will be prepared in advance
and which will be observably present at some moment before the mentioned transaction takes place. 

Then the remaining part of the transaction happens
on the basis of texts that tell a financial story (for instance about mortgage, or about taxation due). These textst 
are the ``real'' means of exchange. 
P did not actually receive the mortgage from the bank as an amount that can be observed at some moment in time, 
and himself made no payment to that amount either. 
P is in fact not sure what he bought by transferring his prepared money either: was it merely 
the right to proceed with the remainder of the transaction,
thus rendering all of it document based? 

P fails to see the sharp distinction between a statement by his employer that guarantees (so it seems) future income,
the valuation of the new home by a chartered surveyor (which the bank seems to take an interest in), the many (unfriendly sounding) 
rules and conditions in the mortgage contract,
the specification of various taxes and fees to be paid for preparing these documents on the one hand,
and the specification (description, calculation) of sums that allegedly are paid as a component of the exchange at large.

\item Carrying the above example somewhat further we notice that  our person P in role SAR
understands perfectly well that asking questions about each step in this complex transaction makes an
unprofessional impression given the strong social foothold of the naive language of buying and selling 
and using money as a means of exchange which
the other participants in this game make use of on a daily basis.
Nevertheless P silently concludes that the deal is primarily if not exclusively 
document based, the more so the higher the mortgage involved, and a that non-trivial and
to some extent also debatable calculation exclusively involving a money of account 
decomposes the transaction and the exchanges that make part of it retrospectively into a price paid 
(thereafter said to be the amount having been
used in exchange) and additional costs (not included in the price that has been paid).

In other words: through systematic use of the money of account function of money a sum is calculated which
inn hindsight is said to have been paid. Money as a means of exchange has been mediated by money as a means of account.

\item Technically P needs to be able to perform, read and audit calculations and tabular clusters of calculations, involving expressions 
having FE as their dimension. 
Of course many other dimensions (time, weight, volume) occur just as well, but the key identities to be checked 
have FE as their dimension. The scalar of expressions of dimension FE can be considered a Walras numeraire. 
The equilibrium sought is of a different kind: 
coincidence in validation and appraisal for incoming and outgoing texts. 

\item Thus, if an invoice is received in the household, P may study it as a text and consult his 
partner that this makes sense. The partner then performs a physical action by means of which payment results. 
Alternatively P may consult his 
partner that the invoice is problematic and that a complaint must be filed.

\item Both for the sake of contributing to his household and for application in the setting of his job, 
P makes a classification of texts about money that he may need to confront. Here is the big difference
with P's colleagues and friends. 
P needs to design this classification as well as a meta-theory for each of the corresponding documents all by himself.
Time and again the plausibility of documents must be assessed and the only options available to P 
is to assume a classification for the document in 
advance and to develop a theory about correct documents of that class. 
A mismatch will then lead to questions and further investigation.

\item Thus P has given up on the very idea that a comprehensive and complete theory of 
money (albeit abstract in the sense outlined above), 
other than that this is a dimension in a calculational
system with dimensional notation not unlike physics. 
For each type of document P needs to determine the rules of the game, that is the criteria that 
determine correctness. P needs to learn to distinguish minor errors from major errors and when to respond to reception of a problematic text 
with subsequent enquiry. 

\item For P money (FE's) is neither a means of exchange, nor a storage of value, nor a unit of account. 
However, the structure theory of documents 
dealing with expressions having dimension FE is P's major expertise or is gradually becoming 
P's best option for demonstrating competence. 
Success is achieved when (after translating to Euro in a final stage) of a text that P has produced or commented
satisfactory responses are obtained from P's environment. If P's income is increased because of achievement (a hypothesis 
that P is not keen to put forward),
this is proven by the fact that a document about such matters is received privately and subsequently handed 
over (after due validation) to P's partner who expresses appropriate satisfaction.  

\item In some cases texts should not only be correct but the contents of texts may be be a source of satisfaction 
or disappointment themselves. P
needs to know which kind of texts lead to secondary assessment of such a kind. When producing texts he will 
often try to satisfy his superiors. This
may be done for instance by demonstrating a profit rather than a loss, or a higher profit rather than a lower one. 

\item P may simplify his life by developing a kind of data flow theory for FE's. Somehow all documents that P is 
asked to consider or write deal
with flows of FE's through implicit networks and correctness means usually not more than an application of an 
appropriate conservation law to the network
underlying the document. Unfortunately converting the data in a document to a precise network description is far 
from easy due to lack of details in most cases. Design heuristics and guesswork may be needed.

\item Mainly in his job P is confronted with the complexity of systems used for bookkeeping and management accounting. 
P needs to understand why
so few people have a thorough understanding of these systems and how to reconcile that observation with the 
phenomenal influence that texts about money
have on operational matters as well as on decisions about future ambitions and organization. 

Needless to say these texts all are based somehow on data 
obtained as ``information'' from the prevailing accounting tools. P needs to incorporate his observations about 
the bookkeeping practices of his employer
consistently with his vision of finance as outlined above. P's ability to do so is quite limited. He has put forward 
some hypothetical explanations of his observations.
\begin{enumerate}
\item Most regular users (and their superiors) of the corporate accounting system don't think in terms of a 
software system that 
needs a definite specification to be 
understood before use.
\item Most if not all users (and their superiors) have no conception of what, in general, a mistake or error of 
either the system or of its usage might 
be. Which 
specifications are violated when a certain error occurs.
\item Most users (and their superiors) don't know that understanding the correctness of the financial support 
system may be a very difficult task, if the 
comparison with computer software is to be trusted.
\item Like in computer software  the number of persons (within a given organization) with intimate knowledge 
of a system (used by that organization)
may be very limited and their worries are likely to be overheard. Exposing their problems and worries is 
potentially unhelpful for their career. Explaining
all relevant details to anybody but their nearest colleagues introduces the combined risk of being perceived 
as unable to communicate at a useful 
level of abstraction and giving away vital information that may lead to one's own role becoming redundant.
\item Comparable to the case of computer software, where the proverbial programmer is considered an 
academic of mediocre importance mainly able
to carry out the rather technical commands obtained from a talented architect or designer, in the case of 
bookkeeping most personnel
is made to believe that their task is of a simple nature in comparison to the core competence of the 
organization for which he works. In both 
cases this is unwarranted. Getting computer programs correct is an extremely difficult task, and the 
same can be said for convincing and transparent
bookkeeping, management accounting and corresponding financial management even for small organizations.
\end{enumerate}

\end{enumerate}

\subsection{Relevance of FMiM\&F for role dependent views on money}
We have outlined in detail a naive theory of money for the SAR role. Such expositions can be imagined for 
other roles of our listing as well as for 
roles that we have ignored. The SAR role is specific because it has to deal explicitly with the phenomena of 
bookkeeping and management 
accounting. Both higher and lower roles may attempt to define themselves in such a way that far less 
intimate knowledge of these 
complicated matters is required or is even useful. For higher roles in terms of the managerial hierarchy an 
additional awareness of the 
economic role of money is required instead of the mechanical and formalistic understanding that we have 
put forward as being  a rational 
thing for some P in role SAR.

We are confident to state that FMiM\&F provides some perspective on 
money that
may be of an independent value even in the absence of working out further details and ramifications.
\begin{enumerate}
\item Money is characterized by a family of IDBR's each of which may be conceived as weighted 
combinations of the ingredients
mentioned in Section \ref{IDBR4M}.
\item For each specific IDBR thus obtained a family of different monies emerges together constituting a moneyage. 
These monies are
amenable to LSCD definitions and together and in combination with related mechanisms such as banks, 
credit card companies,
cashpoints and SCFD's can be developed.
\item Once an SCFD for a moneyage is chosen aspects of an IGUA and IGVA can be selected together potentially
conveying a fairly complete picture.
\item All of this is at its simplest with a formalist view where money's intended role is merely a unit of account. 
More specifically, the formalmoney that is made use of only covers a money of account functionality at the 
expense of all other functions. This
perspective underlies the theory for money for the SAR role  that has been outlined above. A person P in the SAR 
role may adopt that
view while being unable to analyze financial texts convincingly. P may hold that a plurality of classification 
theories for 
financial texts may exists (each of relevance in different circumstances) and that he need not develop one for himself.
\item Therefore P in role SAR may take a just in time approach to the design of the theory that supports classification and 
analysis of financial 
texts. He will do so slowly, and in a call by need fashion (lazy development of theory refinement), when confronted with some 
context that makes an appeal on 
P's understanding of money.
P justifies this viewpoint because he believes that in fact circumstances can be so different that no universal common 
ground can be found, and for that reason he will not search for it either. Rather than looking for similarities between the way
texts about money are understood in very different contexts he approaches each particular context as very specific allowing 
himself to miss the fact that some mechanisms appear just as well and in similar ways in other contexts.
\item One may hold that P's philosophy is needlessly eclectic and that if only P worked harder he would understand that the way
to analyze and understand texts about money is both universal and systematic. But P does not believe this. He feels that all texts
about money are used to convey more managerial (political)  decisions and choices, with quite different motives,
 often portrayed as management objectives forced upon
one by financial circumstances. So P argues that financial texts should be understood and analyzed together with
the management decisions that are grounded on interpretations of those same texts. 
\end{enumerate}

The main difficulty for P in role SAR is to understand how states of affairs as depicted 
in financial texts can give rise to management 
decisions. This seems not to be a matter of deductive logic and finding out how that works is the 
critical competence P needs to acquire. 

The provisional theory for SAR is unfinished regarding the techniques for analyzing texts on money and finance. 
In particular it should be refined with methods for  classification and validation of financial texts.
We consider the development of the SAR theory of money a significant objective of further research. This further
development is context dependent, it may be highly specific for an individual organization from which is working.
Nevertheless the major task is then to develop
the required theory of text classification and analysis  in such a way that it can be used by a range of persons in similar circumstances. 
By turning the further development of SAR's theory of money explicitly into a research question, the unclarity of whether 
or not this theory should or can be shared by other persons in the SAR role disappears. Indeed it can only be further developed 
by means of systematic work if it is acknowledged that
schemes and methods for text classification and analysis thus obtained are not private for any P of role SAR (or of any other role).

\newpage
\section{Virtual money in semi-autonomous organizations}
In this section we will examine in some detail the local financial system of an organization (ORG)
and its implications for an organization's perspective on money. 
This is highlighted from the perspective or employee P in role SAR, 
whose main exposure to matters of money is supposed to be through his activities within ORG. 
P sees himself confronted with virtual money, or more 
precisely\footnote{%
Our use of the phrase virtual money may be confusing. Virtual money is not supposed to exist besides real money in a 
digital  part of the world. The term is used with a reference to computing where a virtual machine is a programmed system 
on top of a real machine. The correspondence is imperfect as virtual machines can coexist with non-virtual machines in the 
same network.} 
money within a virtual financial system.

\subsection{Virtual money for a local financial system (LFS)}
The preceding discussion of P in role SAR and his formalist theory of money gives rise to several subsequent questions: we will
list these questions together with answers.
\begin{enumerate}
\item P distinguishes internal money for ORG from external (customary) money outside ORG. P may (or may not)
take a
formalist position towards external money, but in any case he will take a formalist position towards 
ORG's internal money.
\item When is the FMiMT approach useful for P (in role SAR)? We will assume that this advantage is most visible (and needed) for P
when P needs to operate in his job,
assuming that he operates in the middle of an organization with many thousands of employees. We assume that P plays his SAR role 
in a division DIV
of a large organization ORG. This component is semi-autonomous. It can make many independent decisions but at the same time the
component  needs to fit in the working practices of the overall framework
of the large organization.
\item Why will P profit from the deviations that FMiMT allows from classical and ordinary thinking on money and finance? 
This is because the 
bookkeeping system of ORG provides DIV with a virtual financial system (in the sense of \cite{Merton1995}). 
This virtual financial system may be 
implemented on the basis of a true financial system that makes use of cash and a number of bank 
accounts with varying maturities and interest 
payment conditions, but it is virtual in the sense that the accounts of the system are not directly 
managed by a (commercial) bank. We will speak of the local
financial system LFS of ORG. 
LFS's properties are likely to be dominated by the digital and electronic environment (bookkeeping system) 
that ORG uses for its operation. In
a principled sense ORG confronts DIV and its employees like P with a virtual financial system and 
SAR's formalist perspective on money has been
prepared in order to take care of the hardly predictable idiosyncrasies of such a system.
\item We will write vm(LFS,ORG) for the virtual money of ORG as defined implicitly by LFS. If the influence of a specific
division DIV is significant and different divisions may entertain different virtual monies, we write vm(LFS,ORG)@DIV.
What is wrong with an identification of vm(LFS,ORG) 
in which DIV must operate with external money according to the state's financial system?
We hold that ORG's impact on the LFS can be so thorough that the criteria for 
moneyness are severely compromised inside LFS. Here are some 
mechanisms that may be at work to that effect.

\begin{itemize}
\item ORG may impose in DIV
unpredictable fluctuations of prices that render the unit of account function of vm(LFS, ORG) less useful.
\item ORG may prevent DIV and its employees to make use of amounts that have been 
accumulated on various accounts, thus compromising
the store of value function of `money' on these accounts. 
This may also negatively impact the usage of vm(LFS, ORG) as a means to (standard for) deferred payment.
\item ORG may restrict the transactions that can be made with help of vm(LFS,ORG), thus either compromising the means of 
exchange function or the degree to which availability of money assures freedom of choice.
\item ORG may by combining measures as mentioned above discourage the use of vm(LFS,ORG) 
as a standard of value.
\item vm(LFS,ORG) may be very context sensitive because its usefulness may be
very dependent on the accounts on which it resides. Then using it as a dimension
in a calculus for a money of account becomes untenable. Perhaps each separate account of the LFS 
must be considered an independent dimension.
\item Due to many restrictions vm(LFS,ORG) may not constitute an optimum of liquidity.
\item Although in some way the vm(LFS,ORG) as stored on accounts that 
DIV may access incorporates degrees of freedom (or in other
words freedom of choice for DIV and its employees) the constraints for expressing  
this freedom may be so severe that within DIV no such
freedom is acknowledged or subjectively experienced while the ORG top-management maintains the
existence of that degree of freedom.
\end{itemize}

\item Which differences of analysis may P have to deal with when he compares vm(LFS,ORG) and its implicit reasoning system 
(insofar as P can
figure that out) with the external financial system with the customary reasoning rules used outside ORG?
\begin{itemize}
\item P may find out that if a subunit of ORG performs a task more cheaply than before and 
ends up with virtual money accumulated in some 
account at the end of a period this is hold against him, thus rendering the basis of 
optimization and economization futile or confused.
\item P may find out that the rules of the game are changed so quickly by ORG's top management that he 
feels engaged in a contest where he is by necessity always in a disadvantage. 
The game may turn all conventional reasoning upside down as its only role is to serve as a platform for a 
struggle for power.
\item P may be confronted with almost irrational constraints, 
for instance an ad hoc requirement that some accounts end up positively where other accounts must
end up negatively at the end of some period) which make the virtual money of 
ORG so remote from the external money he knows that
taking external money a heuristic for dealing with the virtual money works against him.
\item P may arrive in the situation which M\"aki describes in \cite{Maki2004} that 
everyone around him (within DIV) acts as if there exists money 
(ORG's virtual money) while unlike M\"aki's judgement he feels compelled to deny that state of affairs.
\end{itemize}
\end{enumerate}

If so many questions can be posed about vm(LFS,ORG)'s status as money, 
why is it meaningful think in terms of virtual money. It is plausible to 
consider vm(LFS,ORG) a virtual money because at the boundaries of ORG its unit 
exchanges at par with the external unit of money and within ORG
no creation or destruction of the amount of vm(LFS,ORG) can take place. 
From the perspective of DIV that is a different matter, however. Policies 
implemented by the ORG management may either impose high taxes (overhead costs)
 or may impose various constraints on the use of accounts.
There is, however, a need for stating virtual moneyness criteria for (candidate) virtual monies. 
It is clear that some logic for dealing with vm(LFS,ORG) is needed.
Thus the combination of familiar conservation laws at the level of ORG 
and the exchange at par at the border justify, formally and intuitively, the
perception of vm(LFS,ORG) as a form of money.

\subsection{The inverse  virtual moneyage preference differential}
In \ref{MoneyagePreference} we have coined moneyage preference in as a measure for the 
commitment of a group of agents to the maintenance of a financial system worth its name. 

In cases where ORG's virtual money deviates from the dominant external money (as a system) 
it is plausible that ORG's management considers
this deviation a useful feature whereas DIV's employees mainly consider that deviation a hindrance. 
That leads to the assumption that the moneyage
preference (regarding ORG's virtual money) has a differential within the hierarchical structure of ORG and
that the moneyage preference increases from top to bottom. 
The more lower one operates inside ORG the more one may profit from an alignment 
between internal and external (virtual and real) money.

It is reasonable to consider vm(LFS,ORG) as degraded money from the perspective of DIV,
if ORG imposes many restrictions on its use. 
That ORG uses its power to force DIV (and other divisions) into the usage of inferior money may be 
considered a consequence of Gresham's law.

\subsubsection{The potential irrationality of virtual money}
When considering an ordinary bookkeeping system for a conventional organization the system of accounts that it provides will,
together with its operations allowing access and manipulation for different stakeholders, constitute a form of  virtual money
(that is a virtual moneyage). This virtual
money satisfies a number of important basic IDBR elements just as much as the underlying money does. For that reason the
moneyness of the virtual money is quite perfect and the moneyness preference 
of most demanding parties is met. However, in the case
that the top-management of an organization has a lower moneyness preference concerning its LFS (local financial system), thus 
probably disregarding the wishes of some lower rank and file the situation can change, and it may even drastically change.

 Virtual money can be made irrational. For instance it could be required that as measured in 
 formaleurocents certain accounts at the end of a period
take a prime number value. At the same time one (DIV financial personnel) may be required to 
predict the number of moves to and from an 
account in well advance. At random times some accounts may be doubled and other accounts 
may be halved. There is no doubt that ORG's 
top-management is able in principle to create a significant deviation of its virtual money from 
any conventional money and to turn that into
an instrument of power. By doing so, vm(LFS,ORG) may acquire unnatural properties.

Should P (in role SAR) complain about this kind of phenomenon if it occurs? We hold that being equipped with a formalist 
position on money P will
be on the lookout for the specificities of virtual monies already if no sign of irrationality is present. P will be fully prepared to 
confront small changes of
the financial system, and to watch for minuscule deviations from ORG's virtual money and the reasoning it supports from what is 
used in general 
in connection with the dominant system of external money.

\subsubsection{Potential rationality requirements for virtual financial systems}
One may only complain about irrationalities of a virtual financial system and its virtual monies if these irrationalities 
can be understood as deviations form a normative view. Merton's \cite{Merton1995} requirements on a financial 
system should be complemented by rationality criteria. 

Besides rationality constraints that may be imposed on a financial system there are other aspects that must be assessed. 
Returning to the case of ORG and DIV, the mechanisms of LFS contribute to the distribution of power within ORG.
DIV may end up with deficits and that may impose a stress on its position within ORG. Cohen \cite{Cohen2005} describes
the two major forms of power that a unit (DIV or in Cohen's case a national state) can apply to deal with such forms of 
stress: the power to delay and the power to deflect. This conceptual scheme is attractive as a tool for analyzing how LFS
enables a clear distribution of power within ORG, for instance by evaluating to what extent both mentioned powers
are kept within limits.

\subsubsection{vm(LFS,ORG)@DIV between two extremes}
P in role SAR must make sense of vm(LFS,ORG)@DIV. In two directions P is 
confronted with
conceptual problems. Developing a theory of external money is not so easy and to use external money as a role model for 
vm(LFS,ORG)@DIV may fail.
On the other hand understanding the details of vm(LFS,ORG)@DIV will have as a prerequisite that LFS is properly understood. 
The latter may be very 
difficult in view of the enormous volume of seemingly unstructured data which it comprises.

In spite of this locus between two impenetrable extremes vm(LFS,ORG)@DIV and its implicit 
reasoning system serve as a platform for decision making
for DIV's management. To analyze how decision making is influenced by the non-standard aspects of 
vm(LFS,ORG)@DIV viewed as money (in spite
of its Gresham's rule compatible degradation)  is a matter left for later work.

\subsection{Research questions amenable to a formal approach}\label{Problems}
We will provide questions that we consider amenable for an approach in 
FMiM\&F style.  For each of these issues it seems to be the case
 that ToM\&F contains rather limited and inconclusive
information and results regarding the question at hand. 

We notice that methods of FMiM\&F are biased towards 
mechanistic and temporal aspects of issues at hand and that this very bias severely limits the 
scope of ambition of that of a formal methods approach.\footnote{%
For instance there is no indication that the question how methods for limiting inflation will impact 
on employment figures can be brought forward via 
FMiM\&F style research.}
But for other important themes such as the impact of highly computerized 
trading with money and stock which are being held in an alternating fashion for the duration of 
a few microseconds only, there is evidence that  FMiM\&F may provide an adequate point of departure.

For SAR an understanding of bookkeeping is of paramount importance.\footnote{%
The story about SAR has an autobiographic flavor: the complexity and seemingly intrinsic obscurity, 
of bookkeeping and management accounting and its impact on all aspects of University
management  has puzzled me for years from the perspective of a computer science head of department, a role that, in the context
of an academic institution in the Netherlands can be rightfully classified as a SAR.}
Questions abound when a formalized understanding of 
bookkeeping and more generally the use of data drive management accounting methods  is sought.
\begin{enumerate}
\item Bookkeeping requires a systematic mathematical and formal treatment. Concepts need to be introduced 
in an incremental way so that different strategies are distinguished at adequate levels. In particular the modular structure of 
bookkeeping needs to be analyzed, that is if different branches of an organization do their bit of bookkeeping in parallel, 
how will things fit together?  
\begin{enumerate}
\item Is there a need for the introduction of a theoretical bookkeeping machine in 
order to model bookkeeping systems in principe? If so, what form will the bookkeeping machine take, and for application in an IT setting
what is a virtual bookkeeping machine.

\item How can one specify the link between an invoice and corresponding subsequent 
payment or payments. This seems to ask for a 
$\pi$-calculus like scheme with the name (identity) of the invoice being bound. 
It should be noticed that the invoice might be circulating amidst
various other system components (agents) before being effected. Finally, however, the issuing part must be able to connect 
an incoming amount to the original invoice. At the highest level of abstraction no more than the very ability to make this
link is required, and this is what $\pi$-calculus can formalize.

\item Is there a useful abstract classification of categories for bookkeeping purposes. In the abstract descriptions of 
two entry bookkeeping that we could find, not even the existence of a classification of various posts is mentioned. That classification,
however seems to be essential for bookkeeping purposes.

\item Financial statements, in particular a balance sheet and a profit and loss statement,  viewed as data types require a 
specification. For such statements modular structure is very important, the modularity requires formalization. Modular structure
can be simultaneous (how do different parts of a statement  for the same time interval fit together) and
sequential (how do different time intervals connect).
\end{enumerate}

\item How can different cost accounting methods be formalized? Are there common mathematical properties for each of those? 
How to specify different methods for compiling a profit and loss statement? Here are asymptotic matters to be investigated: 
what form of soundness does one expect from a profit/loss statement production method? An unsound method systematically yielding 
too high profits will,
sooner or later lead to instability provided the profits are simply divided over shareholders. How to formalize the required
stability properties of cost accounting techniques and the resulting profit loss statements as well as the method for 
profit/loss transfer to subsequent phases.

\item What is a budget? Is it a logical statement, a mere budget expression in a formal notation for budgets,
 is it a forecast, is it a move in a game? 
What relation between budget and actual progress of 
an organization should be asked for. Which kind of tests can make sure that such relations are preserved, or if not that
the problem comes quickly to light.

\end{enumerate}

Each of the above questions may have specialized answers in a particular virtual money, inside a specific organization or division of an organization. We hold that formal methods may be useful for bringing to light commonalities and differences between various
virtual monies and their own traditions of bookkeeping, management accounting, and budget design and defense.

\newpage
\section{Concluding remarks}
Which conclusions can be drawn from the preceding developments (including both appendices). We have no better option than listing
some:
\begin{enumerate}
\item Theories of money can be role dependent. A specific theory of money for the subordinate administrative role (SAR) has been
presented. Another perspective of such a role dependent theory of money is that it determines virtual money, or rather a 
virtual near--money.
\item The inverse moneyage preference differential gives expression to the hypothesis that corporate 
top-management is likely to have a preference to have a virtual money developed within a corporate division 
different from the background money on which that virtual money is  based. 

\item For top-management it may be unproblematic that
a divisional near-money is best not classified as a money, while for divisional functionaries that same situation may be quite frustrating. 
At the same time top-management may not wish to admit the differential, even if it has an advantage from its existence. 
Admission thereof might
trigger an unwelcome  call for a revision of the management accounting system.
\item Imaginative definitions of money can be given. Such definitions produce formalmonies, that is formal counterparts to
monies, rather than monies proper. That gives a rationale to terms like: formalmoney, formalcoin, formaleuro, and formalbitcoin.
The Nakamoto architecture of \cite{BergstraL2013} is an example of a possible definition of a formalbitcoin.
\item Imaginative definitions are developed in stages: IDBR, LSCD, SCFD, SCFD+IGUA, and SCFD+IGUA+IGVA.
\item About the circulation of coins as it takes place in the context of a mainstream money like the Euro, many questions can be posed. Such
questions are better understood in terms of formalmoneys than in terms of the subject money proper. (That is: when theorizing about
hoards of money-items from the Eurocoinage in a hypothetical wallet, one is not speaking of actual coins but rather of their formal counterparts.)
\item With Gesell, Maududi, and Nakamoto we find a chain or line of innovative and non-mainstream financial thinkers 
each having an entirely different perspective on inflation and interests, and on the store of value function that money must fulfil.
From the perspective of attempts to define money, the disparate positions taken by each of these provide a valuable frame of reference.
\item The possibility that the Euro system should not be classified as a money but merely as a near--money
must be taken seriously. That reclassification of the Euro might simplify a search for compatibility with interest free finance. 

Thinking 
along this line it is plausible that proponents of interest free banking will strive for prominence in the context of Bitcoin, 
and it is equally plausible that proponents of interest free banking may be inclined to develop an alternative for 
Bitcoin where purely competitive mining is replaced by a mechanism that profits from some degree of mutual trust among (mining) peers.

\item Bitcoin is innovative as a theory of money by having a complete model of circulation at its core, thereby adding to its non-negative 
rational quantities (best formalized via meadows), logical space (addresses), and spacial distance (needed for security, which is
implicit).
\end{enumerate}

\newpage
\appendix

\section{Dealing with a large volume of prior art}
When writing about money from a perspective of informaticology one encounters a difficulty which merits systematic reflection. 
How to
move from a specialized topic, say X, classified as belonging to a theme P, to a wide open area in a different theme Q? 
Even when armed with insights from P 
the work one performs (prolonging project X now within the arena of Q) might be classified as belonging to Q. Unavoidably 
some
connection with Q needs to be established.\footnote{in the context of this paper: P is TCS, Q is ToM\&F, X is formal methods 
within TCS, and
its prolongation within Q is FMiM\&F.} There is some intersection (meeting point) between P and Q which prospective 
author $a$ may be relatively easily able to spot. The means open to $a$ for obtaining information about Q can be easily listed.
\begin{description}
\item{\emph{Community membership.}} $a$ may be able to find a community active in Q which admits him as a 
member and provides some feedback to his draft papers. This is very nice but it may be difficult to achieve.
\item{\emph{Academic curriculum of Q.}} This takes quite long and $a$ may not be able to do so.
\item{\emph{Consulting reputable researchers from Q.}} This works provided $a$ has access to such persons. 
\item{\emph{Reading books.}} Getting hold of books may be expensive and time consuming.
\item{\emph{Reading (parts of) journal papers in top journals.}} Visiting a library is increasingly outdated.
\item{\emph{Reading (parts of) journal papers in top journals that can be accessed electronically from one's institution.}} 
Now $a$ runs the risk not to notice that his objectives have been pursued by authors who did not 
make it into the top journals of Q.\item{\emph{Consulting electronically available material, including grey literature.}} 
This is cheap and easy.
\end{description}

From these options we have chosen the last one. 
In this section we will specify in some detail a methodology for that way of working.
The methodology can be applied in general, but for readability reasons we have used TCS for P and ToM\&F for Q. 
 
\subsection{A time slice model of available sources}
We consider existing IoM\&F (abbreviated:  IoM\&F$_{2010}$), that is IoM\&F timed at 2010 to be reasonably 
small and for that 
reason the constraint may be formulated that when advancing IoM\&F$_{2011}$ results from IoM\&F$_{2010}$ should be properly 
taken into account. 
As time progresses ($k$ = 2010, 2011, 2012,.. and so on) IoM\&F$_k$ will steadily grow, but it is unlikely that the size of
this area becomes unmanageable before 2020.) 
Assuming that one writes in year k it is reasonable to expect 
that novelty of a result produced  in that year can, with some accuracy, be judged in relation to IoM\&F$_{k-1}$, the 
literature concerning IoM\&F up to year $k-1$. Proper reference to such works 
needs to be made. Only if one has knowledge of active authors and groups a closer match with 
nearly simultaneous work in the same year (preceding months, weeks days or even hours) can be expected and 
should be considered a reasonable and convincing requirement on solid work.

\subsection{How to approach ToM\&F/IoM\&F, a generic approach}
Dealing with ToM\&F/IoM\&F is a far more difficult matter because ToM\&F$_{2010}$ is monumentally large. 
Apart from constituting a wonderful source of ideas ToM\&F$_{2010}$ also presents a 
formidable challenge given its complexity and size 
 if one intends to determine the novelty of a new result (question, viewpoint, argument) $f$ from ToM\&F$_{2010}$.
To begin with we provide a number of rules of engagement. 
Let us assume that one is working in year $k$ (with $k> $ 2009) then ToM\&F$_k$/IoM\&F$_k$ needs to be grasped so that the novelty of 
a result $r$  by author $a$ can be reliably determined.
We may safely assume that author (or group of authors) $a$ in year $k$ may have become acquainted with a subset 
AL$_k^a$ of  ToM\&F$_k$/IoM\&F$_k$. A subset SAL$^a_k$ (for: systematically acquainted literature) may have 
been studied by $a$ in significant detail and is assumed to have been properly understood (by $a$). 
This subset of ToM\&F$_k$/IoM\&F$_k$ need not have been 
accumulated in a digital fashion. Further it may be the case that $a$ has performed key word based searches 
and that SAL$^a_k$ items thus found have been scanned on the presence of specific 
types of content only. SAL$^a_k$ may contain parts of works rather than entire works in some cases.

Author $a$ should at any time be aware of the validity of his own novelty claims (also past ones) in the 
light of SAL$^a_k$. We will assume that result $f$ which $a$ contemplates for inclusion in a paper in year $k$
is either a definition, a theorem (with proof), a conjecture (like a theorem but without proof), or a definition of a for
M\&F research relevant and meaningful concept.

Sufficient criteria for referencing prior art in ToM\&F$_{k-1}$/IoM\&F$_{k-1}$ . 
We suggest that this needs to be done at least under the following conditions:
\begin{enumerate}
\item If $f$ is a definition and a similar definition originally occurs in a paper in ToM\&F$_{k-1}$/IoM\&F$_{k-1}$. Here similarity 
may depend of a transformation, or renaming of the terminology used the paper. 
Such a transformation may be difficult to design, or it may even be controversial.

\item If $f$ is a definition and a rather different definition of the same concept (which again may involve a renaming of terms)
originally occurs in a paper in ToM\&F$_{k-1}$/IoM\&F$_{k-1}$. Here again an assessment of (dis)similarity 
may depend of a transformation of terminology, and differences in terminology may not always be taken for differences in content.
\item If $f$ is a conjecture (or a question) which originally occurs in a paper in ToM\&F$_{k-1}$/IoM\&F$_{k-1}$. Here again 
an assessment of similarity 
may depend of a transformation of terminology
\item If $f$ is a theorem with proof  which originally occurs in a paper in ToM\&F$_{k-1}$/IoM\&F$_{k-1}$. Here once more 
an assessment of similarity 
may depend of a transformation of terminology, furthermore 
methods of proof should be accepted from a different tradition. 
\item If $f$ is a theorem with proof  which can be found by combining at most two results which originally occur in at most 
two (but not more)
different  papers in ToM\&F$_{k-1}$/IoM\&F$_{k-1}$. (modulo a change of terminology) for which
methods of proof should be accepted from a different tradition (or even from different traditions). If three or more sources from
ToM\&F$_{k-1}$/IoM\&F$_{k-1}$ need to be consulted an their results properly combined, 
it is reasonable, and acceptable not to count the combination of three such papers each as prior art. 
(This introduces a domain specific but nevertheless arbitrary threshold on chain length.)
\end{enumerate}

\subsubsection{Claiming novelty} 
Novelty of a work can be acknowledged whenever its result has not been stated before in a very similar form. 
When claiming novelty makes little sense, making reference to prior art is redundant while needlessly 
detracting reader's attention. Less
obvious cases require a choice to be made: if no reference is provided, implicitly a claim to novelty is put forward. In some cases
an author could not care less about novelty or priority of part of his findings, while in an other occasion
a claim concerning the novelty of a result $f$ is made explicitly (which of course must be done honestly, 
should be done properly, and may still be mistaken or even controversial). 

\subsubsection{Declaring a paper obsolete}
Once all explicit novelty claims contained in a paper turn out
to have been defeated by 
subsequent detection of relevant prior art a paper has become obsolete. Making explicit mention of that fact about one's own papers
should be encouraged rather than considered a sign of weakness. Thus one may write that at some date one has concluded that
a paper is obsolete. One may say that the paper has been declared obsolete at date $d$, by its (or another) author(s) $a$, in paper $p$
and so on. Of course in such a case the paper was destined to become obsolete at the date of its writing already, though that fact may 
not have been
known to the author at that time, in spite of an appropriate method of working. 

When writing about a paper $p$ (by author $a$) one may for instance state that it had great impact but surprisingly it was declared 
obsolete three years later 
by author $b$. It is conceivable that subsequently $a$ has contested $b$'s judgement on this matter in his paper $q$, 
and if those arguments are convincing, the state of affairs is that the paper should be  considered having been only partially 
declared obsolete (or even not at all). 

\subsubsection{Considering a paper superseded}
A paper can be superseded 
by one or more subsequent papers 
(from the same or other authors) if that (those) offer(s) improved results while covering at least the same ground. 
Having been superseded 
may or may not be formally declared. In any case this status is quite different (and in fact independent) from the status of having 
been declared obsolete. Most papers become superseded after some time, as this is the rule rather than an
exception in active areas of research, while many papers will never be declared obsolete. It is a reasonable ambition for an author to 
write papers that will never be successfully declared obsolete while being superseded by many forthcoming works preferably 
by other authors.

Novelty claims can 
be stated by means of any form of demarcation of SAL$^a_k$ with the listing of a collection of 
references as an obvious start. Of course for the purpose of describing SAL$^a_k$
one may also refer to all papers in a specific journal, 
the contents of a number of books, all books in a series, the work of one author within some 
period, the work performed in some institution within some period and so on.

Below we list more specific rules concerning the use and acknowledgement 
of prior art which we think should be adhered to in general in circumstances similar to what 
we have mentioned above. 

\subsubsection{How and when to refer to prior art in detail}\label{Prioritycheck}
Below we list some rules concerning referencing to prior art that we think should be adhered to at least intentionally. Actually
complying with such rules is not always easy.
The rules are written in a general form, where X is one's own (or rather author $a$'s) 
line of 
research with progressions X$_k$ (containing the results accumulated until and including year $k$) and where a 
body of relevant literature 
BRL$^a_k$ needs to be taken into 
account in principle. Underlying these suggestions is  the understanding that reading all of BRL$^a_k$ (that is making sure 
that  BRL$^a_k \subseteq$ SAL$^a_k$) is unfeasible.

 Suppose that in $X$ a result (including proposed definitions as well as proposed research questions) $f$ is developed 
 in year $k$. And suppose that references to previous work are considered, in which case should reference 
 and credit be given:
	\begin{enumerate}
	\item A similar but weaker result should be acknowledged. The temporally first reference found should be given, 
		unless the same author 
		has provided a more recent treatment which itself refers his earlier work.
	\item	A similar question/definition should be acknowledged, again with a preference towards oldest sources.
	\item \label{prioritycheck} Whether or not a source is to be mentioned solely depends on the content of the document. 
		Later reception of the source should not enter this choice. 
		Being a member of a community that prefers its own set of standard
		references while ignoring prior art from less known sources is no excuse for lacking references.
	\item Priority (that is: a successful claim on novelty) can only emerge in the limit. Acceptance of a paper by a 
		journal after whatever reviewing or refereeing, however professionally it may have been performed, 
		does not prove any claim on novelty beyond doubt. 
		The claims on novelty should always be made with respect to an explicit grasp of the literature.
	\item	Peer reviewed published work can be ranked somewhat higher than unpublished work. 
		But priority is a matter of time and not a
		matter of (social engineering of and) dealing with reviewing systems. Priority can coexist with independence. 
		Independence (in obtaining a result) can be marked as such and applies if work could not possibly have been
		taking some relevant prior art into consideration. Especially here it is crucial that unpublished so-called preprints 
		or grey literature (more
		often than not produced before a paper has been accepted), need to be taken into account as soon as they could 
		have been obtained. This point of view expresses serious criticism on current journal practice where the fate of
		rejected papers is remarkably unclear.
	\item If at some stage prior art becomes known to an author, this fact need not be viewed as a fiasco per se. Of course that is
		a matter of degree. Indeed systematically searching the literature for prior art and finally, perhaps after many years, 
		spotting a paper which states and proves one's `own' result should rather be considered a virtue, though at the 
		same time it may mark a defeat. 
		Then the prior art thus found must be acknowledged in all forthcoming work which plausibly
		needs to make reference to that result.
	\item If at any time an older reference for a previously obtained result is found that fact needs to be properly recorded. 
		Finding out about the novelty of one's own work is difficult, and doing so for previous work by other 
		authors is of equally difficult in most cases. 
		Therefore at any time one needs to be prepared for a revision of references (in forthcoming work). References made 
		by previous authors cannot be taken for granted as facts about the historic development of research. Of course
		if an author mentions inspiration and information obtained from prior art that fact by itself may be trusted unless
		clear reasons to the contrary emerge. But if it comes to statements of priority which are not clearly based on a 
		sound grip on all prior art the responsibility for the judgements of priority remains with the citing author.
	\item While working in X author $a$ attempts to do new work. While reading BRL$^a_k$ he is both finding inspiration and 
		helpful material for progression within the X-paradigm and making an attempt to refute his own novelty claims. 
		It is impossible in many areas to perform a complete and convincing validation of a novelty claim, by doing a full
		search for its refutation. Thus at some stage results are produced (put on paper, websites, repositories and so on),
		which may not be new after all. Here we may draw from the patenting system. It is reasonable to produce a work that
		contains a result as soon as someone skilled in the art (of doing literature search in the given
		area) with a reasonable investment of 
		time (say one month per 25 
		pages of text written) cannot find a prior work that contains the result or a reasonably close 
		approximation or preceding 
		version thereof. 		
	\item	Of course once a strategy of looking for prior art has been chosen far more specific requirements concerning the 
		quality and comprehensiveness of search 
		should be imposed. If it has become known that authors in some group wrote about a subject then it is reasonable 
		to inspect all of their work. This holds in particular  if finding just one of those works would most likely not even be 
		achieved by a skilled agent
		of literature search (taking a reasonable amount of time for the job, say one month) who has not yet been 
		investing much time in the specific topic at hand.
	\item If the topic of $f$ is discussed by a community  C that is operating (that is writing its papers) within BRL$^a_k$ then
		citation traditions within C need not and even should not be taken as authoritative. Author $a$ may write about C, 
		but need 
		not write as if he is a member of C. C's ideology may or may not be taken for granted. Of course author $a$ may
		intend to catch the attention of members of C. But it is not a rule that this should be attempted nor
		that it shouldn't.
	\end{enumerate}

Armed with these rules of engagement a preparatory investigation of BRL$^a_k$ with a duration of several months 
may suffice for making a credible start with writing about results in the paradigm of $X$. Then there may the
accusation of amateurism, possibly voiced by colleagues from well reputed communities progressively 
contributing to BRL$^a_k$ for many years already, but that is immaterial as long as the rules of engagement are 
followed in a systematic and dedicated fashion. 

\subsubsection{An alternative view}
The rules of dealing with prior art mentioned in this section may be considered misguided or moralistic or simply impractical.
Several ways to compromise these suggestions exist. Here are some alternatives together with some justification, written from
the point of view that ultimately these alternatives are less preferable.
\begin{itemize}

\item When a reference is made to some content it suffices to make use of any published and peer reviewed paper which
contains the needed assertion or quote. It is then left to the discretion of the historians of the field to discuss whether or not
an earlier reference should have been given. 

This protocol is quick, but is is potentially problematic in the hands of so-called communities who can deviate from
rather obvious references
to friendly references by authors whose contribution is a mere copy of original work.  

If speed is vital, this way of working is legitimate as long as it is not done systematically in a series of papers and not
by a community of authors.

\item One can refrain from referencing if an assertion is considered `obvious' or well-known to an extent that referencing
is futile. This is defensible in many cases, but clearly not if the assertion is presented as a contribution which needs to
be defended against actual or conceivable criticism. 

\item One may restrict one's references to `important' prior work which has already been well-received in the scholarly 
literature. This is consistent with not being bothered by finding an original reference. If one refrains from referencing
if no sufficiently `important' reference can be found thus declaring `obvious' what has not been written in `important' 
work, while less prestigious references can be given (and found), one introduces a bias which might in due time be
criticized by authors who advocate a more inclusive perspective on who has been contributing to the relevant field.

\end{itemize}

\subsection{Back to money: informaticological finance}
In this section we briefly survey existing work (by other authors) in informaticological style. 
Work has been done on a number of topics. 
We list the topics but we will not provide a bibliography for each topic. We will mention names of authors 
without giving references.
Using automated search techniques a reader can easily find such references when needed.

\begin{itemize}
\item The most well-known contribution of computer science to finance may be what is now called computational finance. 
Financial market 
simulations, security pricing, investment portfolio analysis, risk analysis, and dealing itself are the core topic of 
computational finance. It has a theoretical branch which constitutes part of IoM\&F.
\item Design and organization of financial data bases, financial spread sheet programs, bookkeeping programs and 
management accounting software. As far as we know this literature in general applies known as well as
experimental techniques for system development to finance oriented problem areas. For that reason no dedicated theory
regarding finance and financial data types has been developed. Most work takes the form of case studies. 

In this area a major part of automation has taken place in the last 50 years. As far as we know at this moment
in terms of theory of money and finance no trace of that work is visible in the research literature. Relational data bases have been 
sufficient to capture financial data and the data types provided for in nearly all program notations have sufficed for
designing systems. Decision support for financial processes has not been singled out from other areas where decision support may
be of use.

\item \label{eCoins} Electronic money, electronic coins, electronic wallets. Many papers have been written about schemes that 
permit some form
of equivalent of coins or banknotes in an electronic way. Such papers specify security requirements, interaction patterns as well as
use cases and then in most cases proceed with a description of cryptographic techniques that allow a desired implication. Important 
work aims at proving implementations of such systems correct and secure.

\item Micro-payments allow small transfers of sums for which ordinary cash payment is useless. This is often done in the context of 
e-commerce applications. Work is similar to the work on electronic money but usually there is an online credit system or debit 
system in place which links transaction to the accounts of parties in these interactions. This work involves 
communication protocols, security protocols, cryptography, testing verification, formal specification, model checking.

\item Ordinary payments in an online setting have also given rise to a diverse literature 
with similar characteristics to the previous items.
\end{itemize}

\subsection{Taking the institutional position of authors into account}
Some hold that `research is what universities do'. This viewpoint, combined with recent quality control systems provides a 
context where some authors, who have carved out for themselves a well-protected place in the system,
seemingly have very little to prove when their work needs to be put forward and published
as being innovative 
research. At the same time others working from outside positions have a  much harder time for doing so. 
Regarding this matter  some additional remarks (observations) are in order:
\begin{itemize}
\item One might claim that only those agents $a$ who start working, studying, and most importantly reading,
at an early age can get the necessary grip on BRL$^a_k$ which enables
them making a contribution to related areas. 

Against this viewpoint we state that at any age, when observing the rules mentioned above a 
researcher may start working in a line of research (which constitutes an X in the preceding discussion) 
which requires him to work according to these guidelines with respect to a large
and often classical body of knowledge that far outweighs what he can ever read in the remaining years still ahead. 

Perhaps Heidegger may be used for this point of view: scientific ``Dasein" asks for the explicit awareness that only limited 
time lies ahead.  Nevertheless one may perceive an
incentive for performing scholarly research irrespective of age and  this `invitation to work' 
may stand in contradiction to the need for adequate reception of existing work. When time passes an author may be increasingly
excused for not having found the relevant references, given the fact that time may run out for doing so while a result $f$ that
he definitely expects to be novel is waiting to be put on paper in an adequate form leaving the task to decide novelty to
other workers whose time is not yet running out. This protocol allows all active authors access to adequate writing.

\item All documented results
of research are to be judged by themselves. Whether or not an author has a reputable academic affiliation is 
of marginal interest only as it comes to evaluation of his work. The same holds in principle for his reputation based on other 
work. In some cases a negative reputation may constitute a valid reason to ignore a paper. It is unreasonable to produce
a sequence of fake documents and then all of a sudden a real one and to ask one's colleagues to be systematically on the 
lookout for the one document unexpectedly containing a significant contribution. 

\item In particular in the field of money and finance some famous people have succeeded from outside positions. Karl Marx
wrote \emph{Das Kapital} in times of unimaginable poverty for current standards, 
hardly supported by a useful affiliation. Georg Simmel, 
who mainly cited Marx in his \emph{Philosophie de Geldes} (1900) only acquired a paid academic 
position in 1916 (aged 58). To date his
work on the psychology of money as well as on the philosophy of money counts as a landmark achievement in sociology (rather
than in philosophy). Although a very prolific and internationally highly visible
writer he was unable to convince his contemporaries of his academic qualifications for
an amazingly long period. When he finally succeeded the great war soon put and end to the institutional support for
his academic activities.

\item It might be required that an outsider upon entering a new area should socialize effectively with renowned 
experts in the field.
Although desirable that mechanism is not easily available, however. In practice it boils down to active community 
membership which is 
hard to achieve in short notice and also hard to obtain from outside a standing research tradition. 
\end{itemize}

\subsection{A survey of ToM\&F$_{2010}$/IoM\&F$_{2010}$}
It is a challenge to provide a meaningful account of the literature on money and ficance ToM\&F$_{2010}$ outside IoM\&F$_{2010}$
which has been written in the last 650 years
within a few pages. In appendix   \ref{appC} we made an  attempt to provide some structure to this body of literature. 
The objective is that this information should be helpful 
to determine whether or not questions posed and statements made in the paper below are new from the perspective of ToM\&F.  
The survey in Appendix \ref{appC} decomposes 
ToM\&F into a number of topical subareas. 
Now a result (question, definition) is considered new if it fails to have been developed within any of 
the mentioned subareas. It is assumed that searching for a particular kind of result is made simpler by having this decomposition at hand. 
Of course the listing of themes of ToM\&F may very well be incomplete. In subsequent versions of this paper that can be improved.

\newpage
\section{A survey of ToM\&F$_{2010}$/IoM\&F$_{2010}$} \label{appC}
This survey serves as a quick scan of the available work. It has been produced exclusively from works 
accessible electronically through the library system of the University of Amsterdam. Books have been left out of consideration.

The literature has been structured in relation to forms of money and institutional embeddings of money. 
Many other ways to decompose ToM\&F are conceivable. 
No attempt has been made to experiment with alternative ways to disassemble the body of literature at hand. 

\subsection{Definition, history, and anthropology of money}
We prefer to use the phrase specification of money rather than definition because most definitions found in the literature are IDBR
definitions which are best viewed as loose specifications rather than definitions.
 
\subsubsection{Definition of money}
Many papers analyze the essence of money in some way or another. Definitions are like ``oxygen is what powers human muscles''.
Fairly imprecise and usually devoid of mechanical content. Such definitions are IDBR definitions in the terminology that 
we have developed above. 

An alternative view of the descriptions of money found in the literature is that these serve as loose
specifications, that is specifications which admit many functionally different realizations.
A specification can be used to recognize a number of features in a practical
context in order to perform a classification of some coherent class of mechanisms. 
A definition, in addition, introduces a concept in such a way that independent reasoning
(and theory development about that concept) is enabled, without any additional recourse to 
unmentioned features of the concept. In \cite{BergstraMiddelburg2010} this
essential property of a definition is referred to as bareness.

We are ultimately interested in definitions of money rather than specifications of money, 
in spite of the fact that this distinction plays no role 
whatsoever in the vast body of ToM\&F. 
With that perspective in mind some of the aspects mentioned below take preference over others. 
For instance both the history and anthropology of money are of lesser importance for that ambition than 
philosophy and sociology of money. 

If the question `what is money' is considered as a foundational issue, many perspectives are possible. 
Often it is assumed that the three first functions listed in Section \ref{IDBR4M}
should apply simultaneously while  separatist viewpoints insists that one or more of the functions can be deleted. 

Philosophical literature adds more aspects to money, in particular to MoE-M and SoV-M, such as the expression of personal freedom,
the development of individual attitudes, an incentive to a development of greed, a sense of national unity or of national identity, 
or even of national pride and or strength. The concept of value is considered in various papers from many angles: psychology, 
economy, philosophy, and law.

Underlying any economically based concept of money seems to be a concept of value. Notions of value have been studied in this 
context of course (George Simmel, Simon Newcomb).

\subsubsection{History of money}
For each feature of the mechanism of money as perceived in any historical episode (including the current one) 
the question can be posed on how and why it came about. Many papers about such matters can be found. 
Indeed for each feature of money an evolutionary 
perspective on its (conceivable) development can be developed and its match with 
historic and economic data can be investigated. This line of work starts with 
Carl Menger (1892). Few of these explanations are conclusive. 
We mention only a single event in the history of money: 1971: president Nixon
terminates the rigid specification of the value of a USA dollar in terms of a fixed amount of gold. 
Metallism has come to an end, at least temporarily.

Burns (1927)  investigated the emergence of ancient money, and so did many other historians of coinage. Usher (1943) and
De Roover (1974) analyze the history of banks. 
Banks are institutions where owners can deposit money either warehoused or against interest 
while its use is entrusted to the banker. 
Such deposits or parts thereof can be transferred to other parties, which requires an (possibly oral) agreement
involving three parties (or their attorneys), the dominant way of working from Roman times to 1500 BC,
or otherwise it can be based on written and signed messages (checks, bill of exchange), 
which became increasingly used in Europe having been sporadic until 1500. 
Nowadays such written messages are increasingly replaced by digital 
electronic communications which in turn can take many forms.

An important branch of historical work concerns the analysis of the works and impact of 
famous authors dealing with money and including for instance:
Plato, Aristotle, Pacioli, Hume, Ricardo, Say, Tooke, Smith, Mill, Marx, Jevons, Menger, Macleod, Wicksell, 
Simmel, Walras, Knapp, Marshall, Fisher, 
Innes, von Mises, Keynes, Schumpeter, 
Hicks, Kaldor, Hayek, Patinkin, Minsky, Robinson, Friedman, Schwartz,
Black, Fama, Kiyotaki, Wright, Ingham, Wray.

Ignoring the demarcation line with IoM\&F, Chaum who pioneered digital cash might as well be added to this list.

Mentioning a name in this listing does not imply that we have ourselves performed any check that their main points of view were first 
promoted by exactly these authors. When mentioning such a point of view together with a concrete reference the
need for such such a check would indeed be implied
by guideline \ref{prioritycheck} of Section \ref{Prioritycheck}.

\subsubsection{Conjectural history of money}\label{ConjHist}
A remarkable style of presentation concerns conjectural history. 
That is about how certain market mechanisms might have come about, 
with the primary objective to improve insight in the mechanism itself rather than its actual history. 
An example is Selgin and White's 1985 description 
of the evolution of a free banking system. 
They position Menger's proposal for the evolution of money as a conjectural history (Menger himself 
may have been more convinced of its historic accuracy, however).
Weir's 2007 production theory of money (\cite{Weir2007}) is a recent example of conjectural history of money.

\subsubsection{Anthropology of money and finance}
A significant literature deals with money systems that have not developed in the western tradition. 
We mention Lapavistas, Ingham, Bohannan, (Bill) Maurer. 
Cowrie (shell) money plays an important role because of its once wide geographic coverage. 
Amstrong (Rossel Island) provides anthropological  information about a very non-trivial form of money which
is beyond mainstream imagination.
Anthropological work is helpful for criticizing mainstream results for instance
predictions that come about from the quantity theory of money. 
Of course the very concept of mainstream ToM\&F is in itself an anthropological 
theme and it its use here and below should be considered with some reservation.

\subsection{Bookkeeping, management accounting, and auditing}
Bookkeeping is understudied in economics as far as we can see. Management accounting and auditing are always explained from
a business perspective, an not from the perspective of what it may add to what one can say about money, 
or with a focus on what is required from an underlying concept of money in order to fit explanations of auditing and accounting.
\subsubsection{Double entry bookkeeping}
	Pacioli (1494) describes so-called double entry bookkeeping. That might be viewed as the start of writing on finance in a 
	western tradition. Double entry bookkeeping is of vital importance, though authors disagree about the precise role it has played. For 
	our purposes the following can be noticed:
	\begin{enumerate}
		\item	It is difficult to find clear descriptions of double entry bookkeeping in the research literature. There are many 
				practically oriented books covering the area but the vast research literature on bookkeeping clearly 
				takes for granted
				that readers understand it already. The most extensive work on formalized bookkeeping is by Mossavich. 
				Far less extensive work is on the Pacioli group, and on the use of Feynman diagrams (Fischer and Braun) for 
				specification of money flows, and approaches making use of linear algebra.
				There are no papers that explain how double entry bookkeeping relates to its possible alternatives in terms of 
				advantages and disadvantages except in verbal and descriptive ways.
		\item		Many authors analyze the history of bookkeeping, but never in technical terms. Remarkable issues seem to be open, 
				such as the question whether or not (and if so in what form) single entry bookkeeping has been a predecessor of 
				double entry bookkeeping.
		\item 	There is some literature on the psychology and the sociology of bookkeeping.
		\item 	There are many papers that deal with the production of balance sheets and profit and loss statements for 
				firms (organization) with a hierarchical (modular) structure. However, we found no evidence that the modularity 
				of balance 
				statements as such has been formulated in terms of formal concepts of modularity as known in software technology.
		\item 	Work on bookkeeping takes money for granted as something that can be counted as natural numbers (often in terms
				of the number of cents in some currency; money of account as perspective). 
				
		\begin{itemize}
		\item Bookkeeping and management accounting literature pays no specific attention 
				to coins, banknotes, transferable accounts or other forms of money. 
		\item We found is no use of statistics, no explicit
				mention of interest calculation, no simultaneous use of different currencies with fluctuating relative value.
		\item	We found no work on bookkeeping (or management accounting) that explicitly views money as a money of account only.
				This separatist perspective on money can have simpler foundations, for instance no quantity theory of money applies to 
				money of account.
			\item	 We found no mention of the fact that conventions of double entry bookkeeping and 
			balance sheet design either depend 
				on assumptions or
				that these must be considered assertions about states of affairs expressed in units of account.
		\end{itemize}
	\end{enumerate}

\subsubsection{Management accounting}
	 Management accounting is the container concept for bookkeeping which contains more systematic and methodological work. 	
	 Here are major topics, directly concerning money and finance:
	\begin{enumerate}
		\item	Cost accounting methods: full costing, marginal costing, direct costing. All papers we found are conceptual
		in style, formal definitions are absent, advantages and disadvantages are usually argued in an informal way. Some
		report field studies comparing different costing techniques, then sample sizes are so small that statistical techniques are 
		not employed.
		\item Objectives of cost accounting: forecasting, risk aversion, increased flexibility.
		\item Auditing in relation to cost accounting. International standardization and uniformization of auditing. Auditing theory 
		is not based on definitions or specifications of money. It takes those for granted without further discussion.
		
	\end{enumerate}
\subsection{Money-item types}
The history of money might be depicted as a history of money classes. 
Before coined metal trade has been driven by means of uncoined
metallic objects. Those objects are never treated as money in economic papers. 
But the mechanism is often mentioned. Starting with coins 
as a money-item type there are three kinds of developments: coinages come and go in a long succession, 
each coinage having its own history. 

Coinages
also demonstrate a development within their money class towards more useful systems which are less costly for the issuer. 
Thirdly the institutional
mechanics of production, distribution and retraction of coins has its own long development with 
various degrees of freedom for both banks and the
public being sought or blocked.

Further money-item types are banknotes, bank (deposit) accounts, electronic and informational monies of various kind.
\subsubsection{Coins}
Work on electronic coins (that is electronic money-items) invariably 
takes the form that we described in Section \ref{eCoins} and
reference to classical economic work on coins will not be found. Excluding work on e-coins we find the following classification 
of work on coins and banknotes.
	\begin{enumerate}
	\item	historic work on coins in past civilizations, including contemplations concerning the concept of money in more general
			terms in the period under investigation. There is a striking lack of uniformity in these writings, in the sense that
			negative information is often deemed incorrect by other authors. We notice the firm statement by \cite{Harris2006} that
			coins were not the only means of settlement in the Roman empire. Or the statement that barter economies without
			coins have not existed. In these papers there is never a mathematically formal account of money or of amounts
			of money, of pertinent transfer protocols, or of security measures used. When coins are discussed deeper
			reflection on monetary issues centers on the following aspects:
			\begin{itemize}
			\item		Which historic development provides a plausible picture of how the use of money through coins came about:
					Has it been an invention, or did it develop through some form of evolution from a more primitive 
					economic system based on pure barter?
			\item 	To what extent were coins used as a store of value. Did that role precede the use of coins as a tool for
					exchange, what role was played by minting in this respect. Which authorities minted coins and how was
					the geographical distribution of these coins, as well as their range of validity related to other currency areas.
			\item 	How important were coins as a tool for financial exchange with respect to other means of settlement, how 
					many coins were in circulation. How credible was an identification of money with coins.
			\item 	How important was the amount precise metals in coins. What explains reductions of this amount. Under which
					circumstances would the purchasing power of coins remain unaffected if the state (or another issuing authority)
					made production cheaper. To what extent can some coin systems be considered fiat money. 		
			\item 	Gresham's law 
					and its interpretations, its evidence, and its applications.
			\item 	Inflation as well as economic crises have been observed in Roman times. Such phenomena are described and 
					explained in various phases of economic history.
			\item 	In which periods is it meaningful to distinguish money of account from money of exchange. Which of
					the two  had temporal precedence, what kind of exchange was used when money of account (say sheep),
					would obviously be unhelpful as money of exchange in many transactions.
			\item 	When designing a coinage the values of different coins seem to obey certain regularities. Several hypotheses
					have been considered concerning that topic. 
			\item 	Counterfeiting and counterfeit detection, counting of coins in relation to weighing, analysis of wear.
			
			\end{itemize}
	\item Modern coin systems
		\begin{itemize}
		\item Counting techniques, production methods, usability problems for various user groups (for instance usability for blind users),
		\item Wear of coins, patterns of usage, regional distribution of coins, counting coins that have been lost.
		\item	Bringing new types of coin into circulation. This issue more recently (second half of 20th century)
			in connection with the role of coins in vending machines.
		\item Criteria for designing coinages, coin types, coin type versions, coin type kinds, and coin type kind instances(i.e.e. coins). 
		The use of coins by various age groups.
		\end{itemize}
	\end{enumerate}
	
\subsubsection{Banknotes}
 The use of banknotes coincides with an increase of scale of economies. Monetary issues become invariably
	connected to economic
	topics. Most papers on banknotes (excluding work on production technology) are more economic in style than any work 
	mentioned on coins. The following themes can be distinguished at least:
	\begin{itemize}
		\item Inconvertibility, i.e. the fact that fiat money is not viewed as a claim on (coin) commodity money, 
		is considered an important conceptual achievement that was first made in China. 
		However, Chinese fiat money regularly suffered from inflation and lack of quantity control and loss
		of credibility. 
		Fiat money that encompasses commodity money in a definitive way is considered a phenomenon of the 20th century. 
		While equilibria with commodity money can be understood (by way of conjectural history) as evolution products from a barter 
		economy using microeconomic mechanisms, 
		the role and development of fiat money is considered much more linked to the influence
		of a state which must behave in such a way that the public trust needed for fiat money to play a 
		dominant role comes about and persists.
		\item Gold (silver) standard: temporal lifting of these standards, advantages and disadvantages 
		of a gold standard, a silver standard, and 
		of a combined standard (bimetallism). 
		\item The size of the circulation of money and its relation to prices; inflation and deflation. Hyperinflation, its social
		consequences and its monetary causes. Management of the volume of
		money in relation to economic policies and ambitions.
		\item The political importance of currency areas. Redesign of currencies and currency areas for political purposes. 
		Temporal development of the relative value of different currencies.
		\item Methods for counting paper money and for measuring the distribution of paper money.
		\item Counterfeit prevention for banknotes (color printing techniques, watermarks, coding and numbering schemes).
	\end{itemize}
	
\subsubsection{Checks, magnetic cards}
Traveller's checks (cheques) have been issued from 1772 onwards and 
constitute physical money according to some definitions. Unsigned
checks that have already been paid for are a form of physical money.
The same holds for magnetic cards which have been loaded from a (bank) demand (=deposit) account.

Credit cards are a different case: they clearly provide a medium for exchange 
but one might claim that the money is created during a transaction,
by way of constructing payer's debt with the bank complementary to 
payee's credit rather than that the money is itself being exchanged.

Credit cards do not provide a store of value, and a credit card is based on a unit of account that must exist beforehand.  

\subsubsection{Bank accounts (notational money)}
Once bank accounts (in particular demand or deposit) accounts
and transfers between accounts within a single bank or between accounts held by different 
 banks are in place  coins and banknotes becomes less
essential. If money is merely moved by means of the modification of database entries 
(lines in a ledger), then  some speak of notational money.
Notational money is massive in size in comparison to coins, banknotes, and checks. 
For that reason a large part of ToM\&F pays no attention 
to any other than notational occurrences of money. Recent theory of banking (Tobin, Goodhart) is written in that style.
Many issues and topics can be distinguished.
	\begin{enumerate}
	\item The emergence of chartal money (transferable accounts).
	\item The concepts of credit and debt, as well as tradable credit and debt.
	\item The role of commercial banks as origins of money is linked to their ability to formally credit a client's (borrower's) account
	with a sum without having any physical counterpart available of that some or a part of it, not even an entry in a bookkeeping system.
	\item While bank accounts represent the majority of money from say 1950 onwards the literature is quite vague about which accounts
	constitute money and which accounts do not. A demand account that allows withdrawals at any moment is considered money, whereas a 
	savings account that matures within 5 years is not. Many authors describe this state of affairs. Tobin (1963) concludes that these lines cannot 
	be reliably drawn. 
	\item Methods for defining what is money and what is not, in a technical sense rather than a philosophical one have been focus of research 
	nevertheless (Friedman and Schwartz), including
	methods of counting money in circulation and methods of relating 
	these data to fundamental economic data.
	\item The philosophical question: "what is money" has been posed in many papers and books.  This issue can 
	already be seriously raised
	 concerning ancient times but the matter becomes 
	increasingly sensitive with more and more forms of money being introduced. Many different answers are provided. No unambiguous
	outcome can be observed. Not even a canon of classical literature on the subject seems to have been agreed upon. 
		\item Interest payment, fluctuation of interest rates. Incorporation of default risk when determining interest rates. Precise
			terminology for guarantee mechanisms (Merton and Bodie 1992).
	\item Liquidity preference, opportunity cost of having interest bearing deposits.
	\item Interest calculation, mortgage calculation, insurance against financial risks.
	\item Term structure of interest rates.
	\end{enumerate}

\subsubsection{Bank Reserves}
Bank reserves are deposits held with a central bank. A vast literature analyzes how large bank reserves (plus currency in stock)
must be in comparison to outstanding loans. In counting methods for the quantity of money, coins, banknotes (card based money), deposit accounts
(provided liquid on relatively short notice), and bank reserves, together constitute so-called  extended base money, 
leaving out the deposits one obtains narrow base money.

\subsection{Financial infrastructure}
A long historical development has given rise to an infrastructure of financial institutions which are tightly
 connected to the concepts of money itself.
\subsubsection{Banking theory}
The literature on banking theory is enormous and fraught with controversies. 
Most papers are of an economic style. Papers on banks as a rule do not
pay attention to formal details of the abstract money types used. 
Phrases like deposit account or debt account are considered sufficiently informative 
for readers in spite of the fact that so many variations of such mechanisms can be imagined a occur in practice. 
Issues dealt with include:
	\begin{enumerate}	
	\item Banks: currency versus commodity money
	\item Banks: inside money versus outside money. The gradual development of services.
	\item Banking structure, clearing houses and clearance mechanisms.
	\item Credit and debt (money as a two sided balance operation).
	\item The capital market (secondary money market). Development of interest rates in relation to different markets. 
	\item Banks issuing private monies, mutual recognition of bank issued monies.
	\item Competition and cooperation between banks.
	\item Free banking (history, conceptual definition, risk analysis, ultimate potential).
	\item What are banks good for: which service are they good at providing, core business determination.
	\item Circuit theory (modeling an economy with consumers, firms and banks and their fundamental interrelations).
	\item Bonds and the mathematics of their trade and value.
	\item Endogenous production (annihilation) of money (by individual  banks).
	\item Bankruptcy of banks.
	\end{enumerate}
	
\subsubsection{National (central) banks}

	\begin{enumerate}
	\item Regulation of banking practice by central authorities. Exogenous production (annihilation) of money 
	\item Interbank clearing mechanism.
	\item Control of the quantity of money and of the rate of interest by central banks. 
	\item Government bonds, secondary money market
	\end{enumerate}

\subsubsection{International mechanisms}
	\begin{enumerate}
	\item Systems of national banks,
	\item Connected currencies,
	\item Policies for keeping relative values of currencies stable.
	\item Supranational institutions (Worldbank, IMF, ECB)
	\end{enumerate}
	
\subsubsection{Political economy and global finance}
Political economy is a huge field in which questions concerning money are commonly viewed as an aspect of wider ranging
developments, movements, conflicts of interest, and so on.  The political economy perspective on money
is obviously very limited regarding mechanical and micro-economic matters. 
In spite of that limitation, works in political economy always are about 
money as if their authors are justified in the implicit claim that monetary technology 
hardly matters unless it volume is extremely large. Indeed
political economy needs to take derivatives and the results of financial engineering 
into account because of their sheer size and impact, 
much less because they (may) constitute novel forms of assets, claims and/or liabilities.

Much of political economy of money has a historic flavor, 
and not merely a conjectural historic style (see Paragraph \ref{ConjHist}). 
Here seems to be an important though unsharp demarcation line. 
We hold that neither the prospects of class struggle nor the many variations of absolute monarchies
need to be understood before or in combination with an attempt to understand money any 
more than that the causes of war need to be understood by those who study the technology of guns.

 	\begin{enumerate}
	\item Location theory investigates the emergence of economic centers of gravity. 
	Recent (post-Keynesian) location theory takes the 
	transportation cost of money as well as spatial liquidity preference differentials into account.
	\item The role of national states in connection with money.
	\item The role of national banks/reserve. Who owns the money. 
	How can responsibilities be shared between state and national bank.
	\item How to merge different currencies in to one.
	\item Currencies with a worldwide role and status. What implications does that role have, if any, for the underlying nation(s).
	\item Monetary cooperation between different currencies.
	\item Composed units of account (e.g ECU and private ECU),
	\item Derivatives of financial products. Futures, options on bonds, currency swaps, CDO's and so on.
	\item Economic cycles, financial crises, skewed development, trade surplus phenomena.
	\item Large scale debts and bankruptcies. (``Too big to fail'' issues.)
	\item (IMF) Managed failures of entire countries.
	\end{enumerate}

\subsection{Dilemmas and contrasts concerning money} 
Throughout the literature one observes a number of contrasts sometimes cast as true differences of
opinion, sometimes cast as philosophical extremes which each philosophy of money needs to acknowledge. That means that
none of these dichotomies need to be considered questions to which research activity is to provide (or has
already produced in the past) conclusive answers. These dichotomies create a multidimensional 
conceptual space in which various concepts 
of money and finance can be located. Many positions can be defined, proposed, and defended, or challenged, modified, and subsequently
once more defended and challenged, each producing fully or marginally
different positions on the issues mentioned in previous subsections of this survey.

Listing these positions is a challenge by itself. We mention some: 

metallism, bimetallism, bullionism, Marxism, chartalism, 
neo-chartalism, Austrian economics (Austrian School), monetarism, neo-Marxism, new-Marxism, 
classical economics, neo-classical economics,
Chicago School, Cambridge School, Stockholm School,
Keynesianism, neo-Keynesianism, new-Keynesianism, post-Keynesianism, horizontalism, 
verticalism, structuralism, circuit theory, NME, BFH-School, post-autistic economics.

Each of these schools/positions combines an economic theory with some more or less
specific viewpoints on money and finance. None of these can be
considered exclusively dealing with money.

In addition to schools and named positions there are many named controversies, though these mostly seem to have a local significance
within a part of the literature only.

Each position has been provided with a range of nuances and variations. 
An axiomatic approach might be useful to find a more systematic
representation of these theories of money and finance but no such result could be found. 
	\begin{enumerate}
	\item Invention versus discovery. Has money come about as a (social) invention or as an (economic) discovery of an 
	outcome of evolution. 
	(Related questions: have pure barter economies ever existed, have the different functionalities of money that are nowadays
	common been developed simultaneously or is this grouping a mere historical accident.)
	\item Fiat money versus legal tender (Simmel versus Knapp).
	\item National money versus international money.
	\item Gresham's law (Thomas Gresham, eponym introduced by Henry Dunning McLeod): 
	an unintended side-effect of institutional monetary governance (Selgin) versus a plain tendency towards the use of 
	more cheaply produced coins and banknotes (Jevons) .
	\item Active money versus passive money. Active money to be understood as an exogenous parameter of equilibrium 
	theory (following Walras),
	while passive money obtains its value by way of the very equilibrium mechanism (e.g. strict adherence to a particular 
	metallic standard). (Exogenous money versus endogenous money.)
	\item Natural interest versus interest set by central bank/governmenet/financial governance preferences,
	\item Metallism versus chartalism (Menger versus Knapp, a contrast that appears in many forms).
	\item State money versus accepted means of exchange (Knapp versus Simmel).
	\item Credit theory of money versus monetary theory of  credit (an alternative coined by 
	Schumpeter with a preference for the first
	 option).
	\item Money as an artifact of state regulation (Black, Fama, Hall, Kitson, Meulen), versus money as defined by 
	the state (Knapp).
	\item Is money physical (commodity money at large) or is it merely a logical dimension? (Murad).
	\item Informational money versus physical money,
	\item Inside money versus outside money,
	\item Money viewed as an economic good by itself versus money conceived as no more than a practical tool that 
	fails to have (and need not have) any intrinsic value. (With Say's law as an extreme.)
	\item Money versus (financial) capital versus wealth (net worth). If money is to be defined in a context with financial capital and
	wealth as preexisting concepts, that provides additional options as well as constraints. For instance: wealth is an assessment 
	or quantification
	of a persons possessions and rights; wealth is independent of
	money and it is primarily measured in terms of utility 
	(leaving aside ecological wealth and cultural wealth); financial capital is that part of an agent's wealth which is held
	for the purpose of creating or obtaining future wealth (including financial capital). Money is the most liquid part 
	of financial wealth.
	\item Mathematical analysis versus (conjectural) historical work. In non-mathematical work on money and finance statements
	regarding (conjectural) historic developments abound. This frequency of mentioning historic developments decreases
	if papers make more use of mathematical methods and tools.
	\end{enumerate}

\subsection{More authors on money and finance}
Many authors not mentioned above have contributed to the ToM\&F. Here we mention some further names we came across but 
whose writings did not quite seem to fit in the rest of our story, at the time of writing.\footnote{Inclusion of
author names in this list indicates that either
at least one of their papers or books was consulted or that significant references to their work have been found and consulted, 
and by no means that a full survey of their contribution was obtained.} 
Of course anybody spending time on the subject of money
(including the principles of bookkeeping and management accounting) may produce another list of names. Nevertheless it
provides readers some perspective concerning the information the author has been exposed to when writing this paper.

Tero Auvinen (socially neutral money),
Luca Fantacci, (history of money, in particular Russia),
Adil Manzoor Bakhisi (Islamic credit cards),
Joerg Bibow (reception of Keynes on liquidity preference),
David Baldwin (the power-money analogy),
Mauro Boianovsky (reception of Wicksell),
Gustav Cassel (bankrate and interest rate),
Naomi Caiden (budgeting),
Robert Clower (criticizes and improves upon Patinkin's incorporation of money in Walrasian equilibrium theory),
Romar Correa (circuit theory),
James Peery Cover (what is money: about Friedman and Schwartz),
Tyler Cowen (on forerunners of NME / BFH-system),
Myra Curtis (savings versus inverstments),
Paul Davidson (interpretation of Keynes on money),
David Dequech (ex ante decision making and subjective expected utility), 
Colin Dey (critical etnography in finance),
Francois Divisia (weighted monetary aggregates)
Eladio Febrero (criticizes neo-chartalism from the perpective of circuit theory), 
Stanley Fisher (banking theory)
Giuseppe Fontana (circuit theory),
Alan Freeman, (follows Keynes and Marx),
Silvio Gesell (built in and accelerated inflation for local monies),
Robert L. Greenfield (BFH system),
A. Graziani (circuit theory),
Kevin D. Hoover (reflection on Walras and Fama),
Jan Kregel, (reception of Marx and Keynes).
Randall Krozner (on forerunners of NME / BFH-system),
L.M. Lachmann (uncertainty and liquidity preference),
A.P.Lerner (functional finance),
Michael Linton (LETS),
A.C. Littleton (value concerns economists, valuation matters for accountants), 
David A. Martin (money is neither unit of account nor means of exchange),
Adolph Matz (development of accounting in Germany around 1940),
Arthur W. Margret (on the reception of Walras),
Bennet McCallum (banking theory, deregulation),
Ludwig Mochty (bookkeeping and analytic hierachy process),
Kenneth S. Most (on Sombart's propositions),
Robert A. Mundell (optimum currency areas),
Satushi Nakamoto (Bitcoin, a technically informational money),
Nicolai Nenovsky (methodology of history of money),
Joerg Niehans (international finance),
Charlott Nyman (sociology of money),
Mohammed Obaidullah (Islamic options),
Mark S, Peacock (catallaxy),
A. Piatt Andrew (quantity theory, value of money),
K. Polanyi (anthropology),
Geert Reuten (Marx on money),
Muhammad Nejatullah Siddiqui (indexing as an alternative to interests),
Simon Smelt (sociology of money),
Eric Timoigne, (history of money),
Kenneth White (creditcart use and household bank accounts),
Leland B. Yeager (BFH system),
V. Zelizer (sociology of money), 

\subsection{Alternatives to mainstream international finance}
Assuming that the western financial system sets standards for international conventions on money and
finance alternatives to it are outside the mainstream. Three alternative approaches will be mentioned.
LETS, Islamic finance, and Bitcoin-like informational money. Names of initiating persons are respectively: Gesell, Maududi,
and Nakamoto (generally assumed to be a pseudonym). We provide some additional comments on Islamic Finance and on Bitcoin.

\subsubsection{Gesell, Maududi, and Nakamoto}
Gesell suggested that local communities can profit from money that has a built in incentive for local spending. Locality in space 
is known from all classic monies, while locality in time is obtained if the particular money-items deprecate at a regular rate until no
value is left. One might understand this deprecation as a negative interest rate because it has no effect on the money of account while it
affects the money as a means of storage by discouraging hoarding. Following ideas of Gesell many LETS 
(local exchange and trading) systems have been developed. Around 2010 thousands of such systems may me in operation world wide.

Maududi sucessfully revitalized the project of interest free finance some 80 years ago in an attempt to define 
Islamic finance as a alternative to mainstream finance. 

Nakamoto (\cite{Nakamoto2008}) initiated Bitcoin, an informational money or near-money which may challenge mainstream finance 
becasueof its implicit criticism on how money is technically managed. Interference with money streams, be in inflationary interference by
banks or central banks unexpectedly issuing new money or by other authorities reversing transfers because of legal other concerns
is made very difficult in a money that complies with the so-called  Nakamoto architecture as presented in \cite{BergstraL2013}.

It is reasonable to classify at least in part the descriptions of 
money in the Islamic world under anthropology of money and finance. 
Such a  viewpoint, however, may be considered asymmetric in view of the persistence of 
Islamic (Shariah based) methods in finance. It is difficult to reconcile both traditions of writing because in the Islamic
tradition there seems to be less explicit reflection concerning what constitutes money. In \cite{BM2011} an attempt has been made
to specify exactly what is meant with interest and what is supposed not to be paid and received for that reason. This turns out to
be deceptively difficult.

\subsubsection{Reclassification of Euro as a near--money, and a dual system design}
We are lead lead to the following question: can the Islamic world be mistaken in its classification of the major mainstream currencies
as monies? We think that may be the case: first of all it seems to be problematic (perhaps also when considered
from an Islamic perspective) that from an Islamic perspective 
so many users of Euro's must be judged as acting against the will of God by 
merely receiving and paying interest on debts expressed in Euro. This moral dilemma (can so many people be considered in error on the
sole basis of a classification of the Euro System as a money, where a money must be understood from the 
perspective of some 1400 years ago)  is solved at once if the Euro System
is not classified as a money but merely as a near--money. 
A justification for this reclassification might be found in the observation that in comparison with a millennium ago the protection against
inflation emanating from governments and central banks who need to finance their debts 
by creating new money has become so low that the store of value function of 
the Euro is compromised to the extent that a rationale for its reclassification to a near--money is present.

In \cite{BergstraL2013} a dual (near--)money system is proposed where a conventional mainstream near--money (after reclassification) is combined with Bitguilder, a hypothetical Bitcoin-like money, for which both debt and interest is by design not an option. This
dual system may comply with mainstream needs and Islamic requirements at the same time, without any further necessity to redesign  
conventional financial products into a Shariah compliant version.

\subsubsection{Interest free finance versus interest permissive finance}
The reclassification of mainstream monies as near--monies (and for that reason non-monies) may be considered by some
to constitute an unreasonably
invasive way to obtain compatibility with Islamic finance and the mainstream financial world, assuming that it works at all.

In the absence of such drastic moves a modification of the terminology of mainstream finance to make room for 
interest free methods seems to be in order. Objections against the interest mechanism cannot be refuted on purely intellectual grounds. 
A simple proposal is to distinguish interest free finance (including Islamic finance)  from interest permissive finance (including mainstream finance).

A noticeable  distinction between ``international finance" and  Islamic finance (a distinction used by Zamir Iqbal in 1999) 
is that in the latter tradition much more emphasis is placed on the question which degrees
of freedom (manifestations of free will) are implicit in the possession of money. 
Iqbal also refers to international finance as about conventional 
financial markets, using mainstream theory, providing major asset pricing models and so on. 
So he depicts Islamic finance as lagging behind
in the area of financial engineering, while denoting non-Islamic or potentially non-Islamic features as mainstream, 
modern and so on. 

Doing away with the pejorative connotations of these terms and phrases will be needed, however, and we see no 
alternative to a distinction between Islamic finance and non-Islamic finance. The
adjective non-Islamic is usually deleted by default but a fully general theory must employ such an 
adjective, or it must make use of a semantic equivalent of it such as interest permissive finance, a more neutral description that
may be considered preferable for that reason.

\subsubsection{Further remarks on Islamic finance}
We understand that Shariah compliant money serves as a money of account and as a 
means of exchange in very much the same way as in non-Islamic 
finance.  The major difference comes about when money is considered as a store of value. In that capacity 
an amount of money may be temporarily borrowed to another agent. 
Islamic finance insists that no financial compensation for the owner of the amount
may be asked or received in that case, which, except for some ground fee, is approximately proportional to both 
amount of borrowed money and the duration of the loan.

Now using an instance or a hoard from
a class of money items as a store of value neither dictates nor
prevents the mechanism of interest payment. 
As a consequence we hold that a theory of money may be required to be so general that it accommodates
both systems, each of which feature a wide range of variations in turn. 
Formally speaking one might consider a Shariah compliant moneyage to be embedded 
in a non-Islamic moneyage with limitations on the freedom of contract.\footnote{%
That viewpoint underlies the concept of a reduced product set finance (RPSF) of \cite{BM2011}.} 
Of course mainstream moneyages feature such limitations just as well, as the notions, of payment, gift,
revenue, and so on need to be distinguished for taxation reasons at least. 
One cannot pay for a purchase by means of a gift and so on. Simmel 
(psychology of money, 1889) notices that money has become less valuable because it 
cannot be used to buy a woman anymore. The fact that
some transactions involving money have become being considered illegal has not had a significant impact on the perceived 
generality of the concept of money in the non-Islamic tradition.

Islamic money not only disallows for interest in either direction. It also rejects gambling without material 
economic objectives, that is when
performing a transaction significant clarity must exist on both sides about what is the truly economic 
basis for it. Further different parties involved in a 
transaction ought to have access to information in a symmetric fashion, and finally all parties 
should be free from coercion.

Pure financial gambling on a large scale is considered bad taste and potentially detrimental in 
many non-Islamic countries. The difference with Islamic finance in 
this respect may be  a gradual matter. That buying and selling should be done on the basis 
of free will underlies all mainstream equilibrium theory, in as far as
performing rational behavior can be considered an act of free will. Clearly this aspect will 
not distinguish between both paradigms. That leaves us 
with the key difference regarding the legitimacy of interest on loans and deposits.

Interest is often considered to be the the price of money, but Marx insisted that money has 
value (and no price) whereas all and only commodities have prices. A major
argument for the necessity of positive interest rates is inflation. Only in the presence 
of interests inflation is bearable for those who intend to use 
money as a long term storage of 
value, generally considered one of its important functions. Inflation in its turn has almost magic connotations. It is a risk, but its counterpart, 
deflation is 
considered even worse, as it may bring the whole economy to a standstill. 
High inflation (and corresponding interest rates) creates difficulties that render a capitalist economy dangerously 
disfunctional, whereas very low inflation introduces the risk of deflation as well as stagnation. 
These considerations lead to moderate inflation 
being considered helpful on macroeconomic grounds, even leading to an (seemingly acceptable) upper 
bound on unemployment (NIARU non-inflation generating rate of unemployment),
estimated around $5\%$ and strongly criticized for its detrimental social consequences by 
Vickery (1998) in his exposition of 15 fatal  fallacies of financial fundamentalism.

\subsubsection{Bitcoin and the definition of money}
At the time of writing we cannot assess whether Bitcoin will survive or not. Irrespective of that outcome, an assessment of its
potential impact is
possible along the lines explored in \cite{BergstraL2013} where an analysis of Bitcoin is made in terms of natural kinds in a manner similar
to applications of natural kinds to the theory of biological evolution.

Bitcoin may also have a profound impact on theories of money including definitional options. Conventional monies provide items
that circulate through a population of users. While future circulation is hard to predict, past circulation may be hard to detect.
Bitcoin provides a system where past circulation leaves no room for any doubt or unclarity. The entire past of Bitcoin circulation
is stored in a distributed database, is visible for all observers, and is used permanently  in the definition of future progression.

As a mathematical theory Bitcoin, depicts money as a quantified phenomenon (say measured in non-negative rational numbers),
which incorporates logical locality and separation (public key based address space) with spatial locality and 
separation (explanatory of the assumption that different users can protect their own secret keys). That the definition of a money must incorporate
clear statements on the mathematics of circulation as well as on logical and spatial dimensions, is a very convincing intuition indeed. 
Bitcoin indicates how that integration of aspects into a definition of money can be achieved. Conventional definitions of money that consist of a listing types of money-items plus a political exposition of governance mechanisms 
only may have become outdated at once by Nakamoto's proposal for Bitcoin.

\subsection{Tentative conclusions from this survey}
From this preliminary survey we drew the conclusion that in no topic among the ones listed here contributions have been written that ought to 
be classified as FMiM\&F. There is no indication that in any of these topics even a modest effort has been made to 
formalize the essence of money,
however understood, any further than what is needed for work that has been done in IoM\&F. In particular 
formalization with the objective of advancing
an understanding of money and finance itself seems not to have been attempted at any scale. 
As important work may have been missed in our survey this conclusion can only be a preliminary one. 

\subsubsection{Usefulness for FMiM\&F}
The survey yields some frame of reference concerning what topics arise if money is investigated professionally in philosophy, (political) 
economy, and (management) accounting science. In particular this survey provides a clue on the rich 
and varied occurrence of works that in some way put 
forward the MoA function  of money at center stage. We conclude that this is the best option for FMiM\&F as well. 
Either in a formalistic style (using formalmoney, formaleuros, formalcoins) or in a more realistic style it is 
plausible to consider a well worked out theory of a money of account which 
provides a framework
for understanding the other functions of money. This is true in spite of the fact that in some cases, 
such as getting food from a vending machine in case of emergency, other roles may take priority 
while leaving the accounting role in a marginal position.

In addition we conclude that work in FMiM\&F ought to pay due attention to the literature and authors listed in this survey. 
All conceivable informal short paragraphs about money seem
to have been already written and thoroughly analyzed. In particular remarks that may seem innocent to an author from a  background in logics and computing 
may relate to quite longstanding debates in political economy and should be dealt with due respect for that reason.

\subsubsection{Absence of mechanical analysis}
While scanning ToM\&F we found very little work that pays significant attention to the mechanical aspects of money. For instance the fact that
coins and banknotes are transported by their owners while deposit accounts seem to rest where they are. The importance of the physical appearance 
of currency is taken for granted, probably because of its long history. Similarly the details of demand accounts are explained only in
very general terms if at all in any paper on the theory of money. Returning to coins and banknotes, procedures for their use in practice are not part of ToM\&F.
What exactly constitutes the action of making a payment is left untouched.

In the overwhelming majority of cases technical (mathematical) work takes the form of the presentation of a 
model situation, often without an explicit analysis of its
relation to real world situations together with a number of agents (agent types) and behavior patterns of 
those each with a menu of probability distributions 
for choices that the agents can make. Then using probability theory equilibrium values or dynamic 
properties (e.g. recurring patterns) are demonstrated.
Often such results are considered properties of the real life setting that motivated the exercise. 
It is probably fair to claim that all work of this nature
makes use of formalmoney rather than of real money although this is usually not stated that way. 


\subsubsection{Can one extract a definition of money from by ToM\&F?}
The philosophical question what is money has been posed by many authors and its answer
undergoes a steady evolution. No definition of money 
can do much more than to codify current practice and to reflect about the plurality of services that a certain range of financial
technologies provides. As to the question 
whether or not ToM\&F produces a convincing or even plausible definition of money the situation is hard to 
assess. ToM\&F is written as if money were like music:
it is more or less clear what it is and ontological battles are marginal. 

A comparison between money classes and music instruments seems to be helpful.  From the summer of 2010 
onwards the vuvuzela is a worldwide famous acoustic instrument. Should the question whether or not it is a 
music instrument
be taken seriously. Its use in South Africa during the world championships organized by FIFA  has been 
explained to the public as the combined expression of cultural identity and local tradition, which must
be respected for that very reason, in spite of the health problems that it may cause when used without 
sufficient care. Now the question whether or not the sounds of the vuvuzela are to be classified as music is 
fundamentally unimportant because even if it is perceived as music that changes little to the existing
classification of musical instruments and of traditions of musical expression.

Returning to money: equally alive is quantity theory with some stating that governments 
saving banks  via new (forms of) government debts
 run the risk of causing inflation (as an
application of quantity theory), while others claim that governments refraining from similar 
actions may steer their citizens towards stagnating deflation with 
far worse consequences. In this context, however, the question whether or not some financial 
products are counted as money are very prominent because
the sheer size of their holdings may dwarf that of classical products, even in combination. 
In other words, old money classes (or their value) are not at all immune from the
classification questions of new financial products as monies.

Additional remarks on the definition of money:
\begin{itemize}
\item  The moneyness of (instances of) known money classes is not rendered problematic 
by the introduction of new money classes and new near--monies. 
But the economic value of known monies may change due to the introduction of other monies (or near--monies).
\item Thus the ToM\&F story of classical money (which grows with time, though with some delay) 
and its use as well as it economic role is very stable.
\item At the same time the viewpoint that one knows what money is if one understands the classical 
money classes is as unwarranted as the 
assumption that knowledge of classical instrumental and vocal music in a European tradition 
provides sufficient knowledge of music at large. Today's popular music
is far larger as a social and an economic phenomenon.
\item Comparing new monies (derivatives) to popular music fails when it comes to appreciation. 
The appreciation of new monies may rather
be compared to that of modern music in the classical tradition: 
only very few people have the expertise and experience to adequately appreciate such
forms of music. Perhaps popular music (or rather its more visible manifestations) might 
rather be compared to the multitude of investment 
options, credit card systems, electronic payment devices, webshops, lotteries, retail chain based saving 
and discount mechanisms which are sometimes successfully sold as financial  innovations.
\item Now modern classical music may be considered a marginal niche between 
well-known classical music and well-known popular music. Modern
financial techniques are in terms of public perception perhaps marginal in a 
similar way, as constituents of the world-wide picture of monies they are not.
\item Perhaps these uncertainties  should be taken as additional and even as 
fundamental constituents of the definition of money: the public does not know and 
cannot know exactly what explains its current value (neither can it know what is 
or what should be counted as money at any particular moment in time), 
and it cannot know what threatens its future value. So the
public needs to trust the overall commitment of the political structure to keep 
the financial system afloat in order to trust that its money stock will be 
protected against the unpredictable impact of future financial innovations.
\end{itemize}

\end{document}